\documentclass{article}

\usepackage{microtype}
\usepackage{graphicx}
\usepackage{booktabs} 
\usepackage{bm}
\usepackage{amsthm}
\usepackage{lipsum,multicol}
\usepackage{amsfonts}
\usepackage[mathscr]{euscript}
\usepackage{url}
\usepackage{eepic, epsfig, amsmath, amssymb, amsthm,latexsym, setspace}
\usepackage{rotating}
\usepackage{nicefrac}
\usepackage{lipsum,multicol}
\usepackage[parfill]{parskip}
\usepackage{caption}
\usepackage{subcaption}
\usepackage{float}
\usepackage{algorithm}
\usepackage{algorithmic}
\usepackage{times}
\usepackage{xcolor}
\usepackage{cite}

\usepackage[shortcuts]{extdash}

\newtheorem{theorem}{Theorem}[section]

\newtheorem{lemma}[theorem]{Lemma}

\numberwithin{equation}{section}

\newcommand{\R}{\mathbb{R}}

\newcommand{\x}{\mathbf{x}}
\newcommand{\y}{\mathbf{y}}
\newcommand{\z}{\mathbf{z}}

\newcommand{\e}{\mathbf{e}}

\def\bfb{{\mathbf b}}

\def\bfe{{\mathbf e}}

\def\bfv{{\mathbf v}}
\def\bfw{{\mathbf w}}
\def\bfx{{\mathbf x}}
\def\bfy{{\mathbf y}}
\def\bfz{{\mathbf z}}

\def\bfR{{\mathbf R}}

\def\mcM{{\mathcal M}}

\newcommand{\mbR}{\mathbb{R}}
\newcommand{\mcN}{\mathcal{N}}
\newcommand{\mbP}{\mathbb{P}}

\usepackage{hyperref}

\begin{document}

\title{Outlier Detection using Generative Models with {{Theoretical}} Performance Guarantees\footnote{Jirong Yi and Anh Duc Le contribute equally.}}
\date{\today}
\author{Jirong Yi\thanks{Department of Electrical and Computer Engineering, University of Iowa, Iowa City, USA}
\and Anh Duc Le\thanks{Department of Electrical and Computer Engineering, University of Iowa, Iowa City, USA}
\and Tianming Wang\thanks{{Institute of Computational Engineering and Sciences, University of Texas, Austin, USA}}
\and Xiaodong Wu\thanks{Department of Electrical and Computer Engineering, University of Iowa, Iowa City, USA}
\and Weiyu Xu\thanks{Department of Electrical and Computer Engineering, University of Iowa, Iowa City, USA. Corresponding email: weiyu-xu@uiowa.edu}
}

\maketitle

\begin{abstract}

{{
	This paper considers the problem of recovering signals from compressed measurements contaminated with sparse outliers, which has arisen in many applications. In this paper, we propose a generative model neural network approach for reconstructing the ground truth signals under sparse outliers. We propose an iterative alternating direction method of multipliers (ADMM) algorithm for solving the outlier detection problem via $\ell_1$ norm minimization, and a gradient descent algorithm for solving the outlier detection problem via squared $\ell_1$ norm minimization. We establish the recovery guarantees for reconstruction of signals using generative models in the presence of outliers, and give an upper bound on the number of outliers allowed for recovery. Our results are applicable to both the linear generator neural network and the nonlinear generator neural network with an arbitrary number of layers. We conduct extensive experiments using variational auto-encoder and deep convolutional generative adversarial networks, and the experimental results show that the signals can be successfully reconstructed under outliers using our approach. Our approach outperforms the traditional Lasso and $\ell_2$ minimization approach.

}}

\end{abstract}


\section{Introduction}
\label{Sec:Introduction}

{{

	{{Recovering}} {{signals}} from its compressed measurements, also called compressed sensing (CS), has found applications in many fields, such as image processing, matrix completion and astronomy \cite{CandesRombergTao:2004,Donoho:2006:1,Candes:2009,Lustig:2007,Jaspan:2015,xu_separation-free_2017,cai_robust_2016,bobin_compressed_2008}. In some {{applications}}, faulty sensors and malfunctioning measuring system can give us measurements contaminated with outliers or errors \cite{mitra_analysis_2013,carrillo_robust_2010,studer_recovery_2012,xu_sparse_2013}, {{leading to}} the problem of recovering the true signal and detecting the {{outliers}} from the mixture of them. Specifically, suppose the true signal is $\bfx\in\mathbb{R}^n$, and the measurement matrix is $\mathcal{M}\in\mathbb{R}^{m\times n}$, then the measurement will be
	\begin{align}\label{Defn:MeasurementWithOutlier}
	\y = \mathcal{M}\x + \e + \eta,
	\end{align}
	where $\e\in\mathbb{R}^{m}$ is an $l$-sparse outlier {{vector}} due to sensor failure or system malfunctioning, and the $l$ is usually much {{smaller}} than ${m}$. Our goal is {{to recover}} the true signal $\x$ and detect $\e$ from the measurement $\y$, and we call it the outlier detection problem. {{The outlier detection problem has attracted}} the interests from different fields, such as system identification, channel coding and image and video processing \cite{Bai:2017,xu_sparse_2013,barner_nonlinear_2003,CandesTao:2005,wright_dense_2010}. For example, in the setting of channel coding, we consider a channel which can severely corrupt the signal the transmitter sends. {{The useful information is represented by $\x$ and is encoded by a transformation $\mcM$. When the encoded information goes through the channel, it is corrupted by gross errors introduced by the channel.}}

	There exists a large volume of {{works}} on the outlier detection problem, ranging from the practical algorithms for recovering the true signal and detecting the outliers \cite{CandesTao:2005,wright_dense_2010,xu_sparse_2013,wan_robust_2017,popilka_signal_2007}, to recovery guarantees \cite{CandesTao:2005,wright_dense_2010,xu_sparse_2013,studer_recovery_2012,popilka_signal_2007,mitra_analysis_2013,carrillo_robust_2010,candes_highly_2008}.
	{{
	In \cite{CandesTao:2005}, the channel decoding problem is considered and {{the authors proposed $\ell_{1}$ minimization for {{outlier}} detection. By assuming {{$m>n$}}, {{Candes and Tao}} found an annihilator to transform the error correction problem into a basis pursuit problem, and established the {{theoretical}} recovery guarantees. Later in \cite{wright_dense_2010}, Wright and Ma considered the same problem in a case where $m< n$. By assuming the true signal $\x$ is sparse, they reformulated the problem as an $\ell_1 \backslash \ell_1$ minimization. Based on a set of assumptions on the measuring matrix $\mcM$ and the sparse signals $\x$ and $\e$, they also established {{theoretical}} recovery guarantees. Xu et al. studied the sparse error correction problem with nonlinear measurements in \cite{xu_sparse_2013}, {{and proposed {{an}} iterative $\ell_1$ minimization to detect outliers}}. By using a local linearization technique and an iterative $\ell_1$ minimization algorithm, they showed that with high probability, the true signal could be  recovered to high precision, and the iterative $\ell_1$ minimization algorithm will converge to the true signal.
	}} All these algorithms and recovery guarantees are based on techniques from compressed sensing (CS), and usually {{they assume that the signal $\bfx$ is sparse over some basis, or generated from known physical models}}. In this paper, by applying techniques from deep learning, {{we will solve the outlier detection problem when $\bfx$ is generated from generative model in deep learning}}. {{We}} propose a generative model approach for solving outlier detection problem.

	{{Deep learning \cite{Goodfellow:2016} has attracted attentions}} from {{many}} fields in science and technology. Increasingly more researchers study the signal recovery problem from the deep {{learning}} prospective, such as implementing traditional signal recovery algorithms by deep neural networks \cite{yang_admm-net:_2017,metzler_learned_2017,gupta_cnn-based_2018}, and {{providing}} recovery guarantees for nonconvex optimization {{approaches}} in deep learning \cite{bora_compressed_2017,dhar_modeling_2018,wu_sparse_2018,papyan_theoretical_2018}. In \cite{bora_compressed_2017}, the authors {{considered}} the case where the measurements are contaminated with Gaussian noise {{of}} small magnitude, and they proposed to use a generative model to recover the signal from noisy measurements. {{Without}} requiring sparsity of the true signal, they showed that with high probability, the underlying signal can be recovered with small reconstruction error by $\ell_2$ minimization. However, similar to traditional CS, the techniques presented in \cite{bora_compressed_2017} can perform {{bad}} when the signal is corrupted with gross {{errors}}. Thus it is of interest to study the conditions under which the generative model can be applied to the outlier detection problems.

	In this paper, we propose a framework for solving the outlier detection problem by using techniques from deep learning. Instead of finding directly $\x\in\mathbb{R}^n$ from its compressed measurement $\y$ under the sparsity assumption of $\x$, we will find a signal $\z\in\mathbb{R}^k$ which can be mapped into $\x$ or a small neighborhood of $\x$ by a generator $G(\cdot)$. The generator $G(\cdot)$ is implemented by a neural network, and examples {{include}} the variational auto-encoders or deep convolutional generative adversarial networks \cite{Kingma:2013,Goodfellow:2014,bora_compressed_2017}.

	 {{The generative-model-based}} approach is {{composed}} of two parts, i.e., the training process and the outlier detection process. In the former stage, by providing enough data samples $\{(\z^{(i)},\x^{(i)})\}_{i=1}^N$, the generator will be trained to map $\z^{(i)}\in\mathbb{R}^k$ to $G(\z^{(i)})$ such that $G(\z^{(i)})$ can approximate $\x^{(i)}\in\mathbb{R}^n$ well. In the outlier detection stage, the well-trained generator will be combined with compressed sensing techniques to find a low-dimension representation $\z\in\mathbb{R}^k$ for $\x\in\mathbb{R}^n$, where the $(\z,\x)$ can be {{outside}} the training dataset, but $\z$ and $\x$ has the same intrinsic relation as that between $\z^{(i)}$ and $\x^{(i)}$. In both of the two stages, we make no {{additional}} assumption on the structure of $\x$ and $\z$.

	{{For}} our framework, we first give  necessary and sufficient conditions under which the true signal can be recovered. We consider both the case where the generator is implemented by a linear neural network, and the case where the generator is implemented by a nonlinear neural network. In the {{linear neural network}} case, the generator is implemented by a $d$-layer neural network with random weight matrices $W^{(i)}$ and identity activation functions in each layer. We show that, when the ground true weight matrices $W^{(i)}$ of the generator are random matrices and the generator has already been well-trained, {then with high probability, we can theoretically guarantee successful recovery of the ground truth signal under sparse outliers via $\ell_0$ minimization}. Our results are also applicable to the nonlinear neural networks.
	
	We {{further}} propose an iterative alternating direction method of multipliers algorithm and gradient descent algorithm for the outlier detection. Our results show that our algorithm can outperform both the traditional Lasso and $\ell_2$ minimization approach.

	{{We summarize our contributions in this paper as follows}}:

    \begin{itemize}
	\item We propose a {{generative model based}} approach for the outlier detection, {{ which further connects compressed sensing and deep learning.}}

	\item We establish the recovery guarantees for the {{proposed approach}}, which hold for generators implemented in linear or nonlinear neural networks. Our analytical techniques also hold for deep neural networks {{with arbitrary depth.}} 
	
	\item	Finally, we propose an iterative alternating direction method of multipliers algorithm for solving the outlier detection problem via $\ell_1$ minimization, and a gradient descent algorithm for solving the same problem via squared $\ell_1$ minimization. We conduct extensive experiments to validate our theoretical analysis. Our algorithm can be of independent interest in solving nonlinear inverse problems.

	\end{itemize}

	The rest of the paper is organized as follows. In Section \ref{Sec:ProblemStatement}, we give a formal statement of the outlier detection problem, and give an overview of the framework of the generative-model approach. {{In Section \ref{Sec:ADMMAlgorithm}, we propose the iterative alternating direction method of multipliers algorithm and the gradient descent algorithm. We present the recovery guarantees for generators implemented by both the linear and nonlinear neural networks in Section \ref{Sec:PerformanceAnalysis}. {{Experimental}} results are presented in Section \ref{Sec:NumericalResults}. We conclude this paper in Section \ref{Sec:ConclusionsandFutureDirections}.}}

	{{

	{\bf Notations:} {{In}} this paper, we will denote the $\ell_1$ norm of an vector $\x\in\bfR^n$ by $\|\x\|_1 = \sum_{i=1}^n |\x_i|$, and {{the number of nonzero elements of an vector $\x\in\bfR^n$ by $\|\x\|_0$}}. For an vector $\x\in\bfR^n$ and an index set $K$ with cardinality $|K|$, we {{let}} $(\x)_K\in\bfR^{|K|}$ {{denote}} the vector consisting of elements from $\x$ whose indices are in $K$, and {{we let}} $(x)_{\bar{K}}$ {{denote the}} vector with all the elements from $\x$ whose indices are not in $K$. {{We denote the signature of a permutation $\sigma$ by ${\rm sign}(\sigma)$}}. {{The value of ${\rm sign}(\sigma)$ is $1$ or $-1$}}. The probability of an event $\mathcal{E}$ is denoted by $\mathbb{P}(\mathcal{E})$.

	}}

}}

\section{Problem Statement}\label{Sec:ProblemStatement}
	In this section, we will model the outlier detection problem from the deep learning prospective, and introduce the necessary background needed for us to develop the method based on generative model.

{{
Let generative model $G(\cdot):\mathbb{R}^k\to\mathbb{R}^n$ be implemented by a $d$-layer neural network, and a conceptual diagram for illustrating the generator $G$ is given in Fig. \ref{fig:GeneralNNStructure}. 
\begin{figure}[htb!]
\begin{center}
\includegraphics[width=\textwidth]{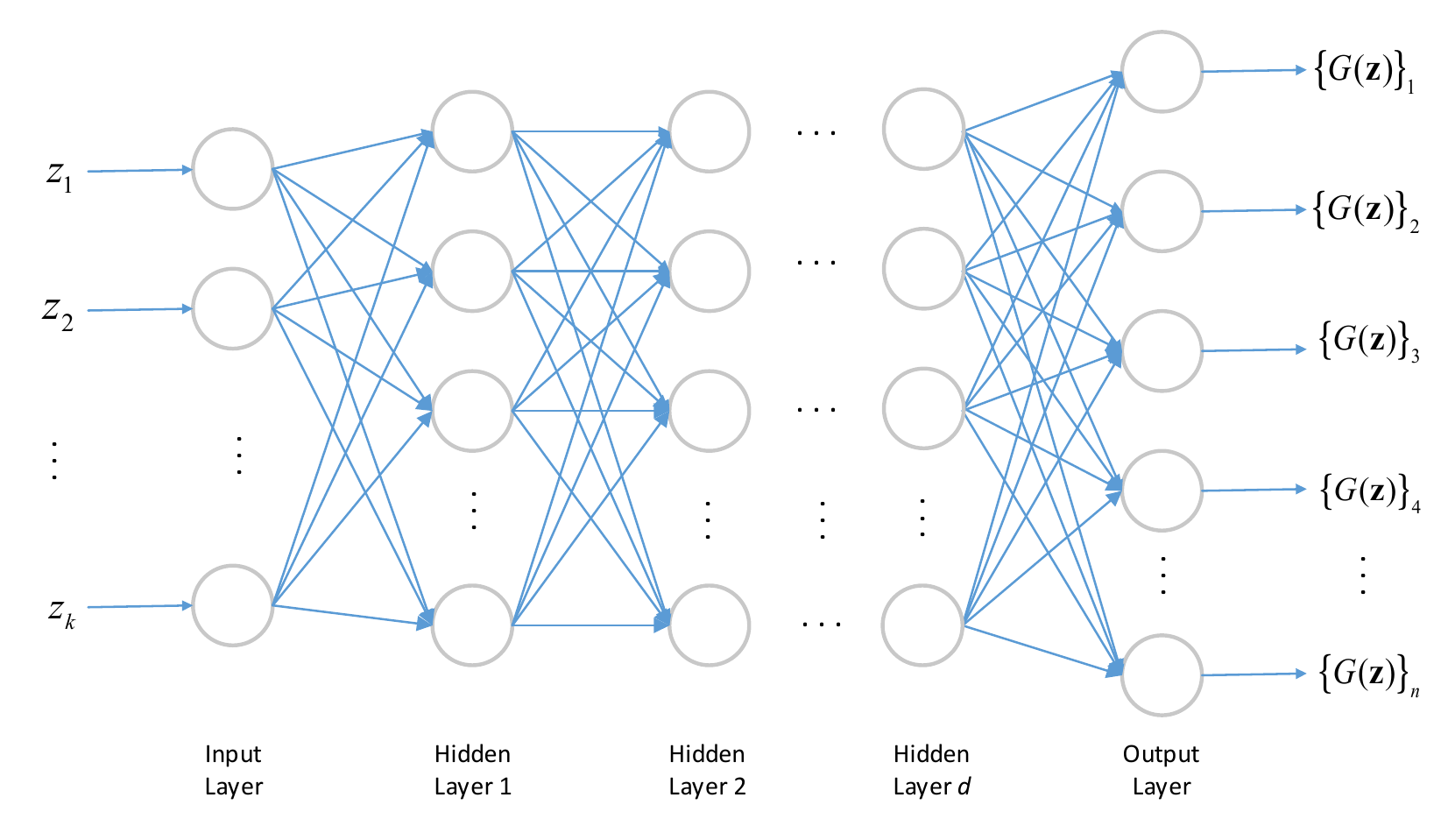}
\caption{General Neural Network Structure of Generator}
\label{fig:GeneralNNStructure}
\end{center}
\end{figure}

Define the bias terms in each layer as
$$
\bfb^{(1)}\in\mbR^{n_{1}}, \bfb^{(2)}\in\mbR^{n_{2}}, \cdots, \bfb^{(d-1)}\in\mbR^{n_{d-1}}, \bfb^{(d)}\in\mbR^{n}. 
$$
Define the weight matrix $W^{(i)}$ in the $i$-th layer as
\begin{align}\label{Defn:WeightMatrix-2}
W^{(1)} = [w^{(1)}_1,w^{(1)}_2,\cdots,w^{(1)}_{n_1}]^T\in\mathbb{R}^{n_1 \times k},
w^{(1)}_j\in\mbR^k, j=1,\cdots,n_1,
\end{align}
\begin{align}\label{Defn:WeightMatrix-1}
W^{(i)}
= [w^{(i)}_1,w^{(i)}_2,\cdots,w^{(i)}_{n_i}]^T\in\mathbb{R}^{n_i \times n_{i-1}},
w^{(i)}_j\in\mbR^{n_{i-1}}, 1< i < d, j=1\cdots,n_i, 
\end{align}
and
\begin{align}\label{Defn:WeightMatrix-3}
W^{(d)} = [w^{(d)}_1,w^{(d)}_2,\cdots,w^{(d)}_{n}]^T\in\mathbb{R}^{n \times n_{d-1}},
w^{(d)}_j \in\mbR^{n_{d-1}}, j=1,\cdots,n.
\end{align}
The element-wise activation functions $a(\cdot)$ in different layers are defined to be the same, and some commonly used activation functions include the identity activation function, i.e.,
\begin{align*}
[a(x)]_i = x_i,
\end{align*}
 ReLU, i.e.,
\begin{align}\label{Defn:ActivationFunctionReLU}
[a(\bfx)]_i =
\begin{cases}
0, {\rm if\ } \bfx_i<0,\\
\bfx_i, {\rm if\ } \bfx_i\geq0,
\end{cases}
\end{align}
and the leaky ReLU, i.e.,
\begin{align}\label{Defn:ActivationFunctionLeakyReLU}
[a(\bfx)]_i =
\begin{cases}
\bfx_i, {\rm if\ } \bfx_i\geq 0,\\
h \bfx_i, {\rm if\ } \bfx_i<0,
\end{cases}
\end{align}
where $h\in(0,1)$ is a constant. Now for a given input $\bfz\in\mathbb{R}^k$, the output will be
\begin{align}\label{Defn:IORelationArbitrayLayerArbitraryActivationFunction}
G(\bfz) = a\left(
      W^{(d)}a\left(
      W^{(d-1)} \cdots a\left(
                         W^{(1)}\bfz + \bfb^{(1)}
                        \right)  \cdots + \bfb^{(d-1)}
             \right)+ \bfb^{(d)}
\right).
\end{align}
}}

Given the measurement vector $y\in\mbR^{m}$, i.e., 
$$
\bfy = \mcM\bfx + \bfe,
$$
where the $\mcM\in\mbR^{m\times n}$ is a measurement matrix, {$\bfx$} is a signal to be recovered, and the $\bfe$ is an outlier vector due to the corruption of measurement process. {The $\bfe$ and $\bfx$ can be recovered via the following $\ell_0$ ``norm'' minimization if $\bfx$ lies within the range of $G$, i.e.,}
$$
\min_{\bfz\in\mbR^k} \|\mcM G(\bfz) - \bfy\|_0.
$$
However, the aforementioned $\ell_0$ ``norm'' minimization is NP hard, thus we relax it to the following $\ell_1$ minimization for recovering $\x$, i.e.,
	    $$
	    \min_{\z\in\mathbb{R}^k} \|\mathcal{M}G(\z) - \y\|_1,
	    $$
	    where $G(\cdot)$ is a well-trained generator. When the generator is well-trained, it can map low-dimension signal $\z\in\mathbb{R}^k$ to high-dimension signal $G(\z)\in\mathbb{R}^n$, and the $G(\z)$ can well characterize the space $\mathbb{R}^n$. These well-trained generators will be used to solve the outlier detection problem, and we give a conceptual flowchart of the outlier detection stage in Fig. \ref{fig:diagram}. Given a truth signal $\x\in\bfR^{n}$, we can get the compressed measurements $\y\in\mathbb{R}^m$ defined in (\ref{Defn:MeasurementWithOutlier}) from $\x$. The goal is finding a $\z\in\bfR^{k}$ which can be mapped into $\hat{\x} = G(\z) \in\bfR^{n}$ satisfying the following property: when $\hat{\x}$ goes through the same measuring scheme, i.e., $\mcM$, the measurement $\hat{\y}\in\bfR^{m}$ will be close to the measurement $\y$ from the truth signal $\x$.  {{Note that in Fig. \ref{fig:diagram}, we do not include the training process.}} We further propose to solve the above $\ell_1$ minimization by alternating direction method of multipliers (ADMM) algorithm which is introduced in Section \ref{Sec:ADMMAlgorithm}. We also propose to solve the outlier detection problem by a gradient descent algorithm via a squared $\ell_1$ minimization, i.e.,
	    $$
	    \min_{\bfz\in\mbR^k} \|\mcM G(\bfz) - \bfy\|^2_1.
	    $$

{{
{\bf Remarks:} In analysis, we focus on recovering signal and detecting outliers. However, our experimental results consider the signal recovery problem with the presence of both outliers and noise. 
}}


\begin{figure}[htb!]
\begin{center}
\includegraphics[width=\textwidth]{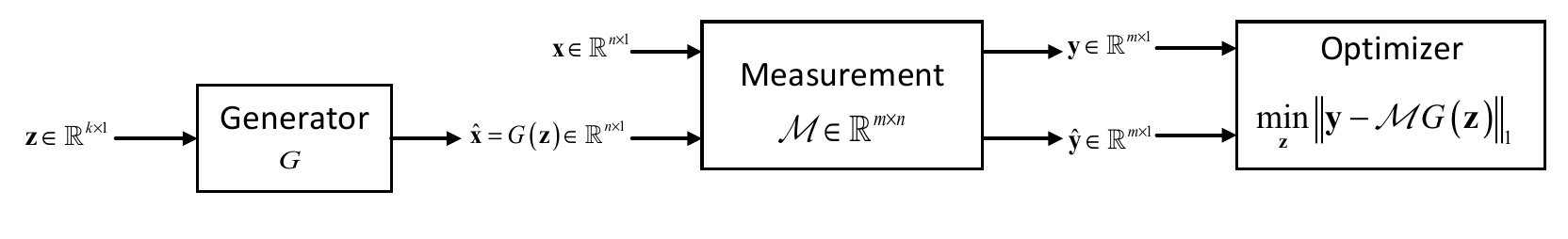}
\caption{Deep Learning Based Compressed sensing system}
\label{fig:diagram}
\end{center}
\end{figure}

{{Note that in \cite{dhar_modeling_2018}, the authors {considered} the following problem
	\begin{align*}
	& \min_{\bfz,\bfv} \|\bfe\|_0 \\ 
	& {\rm s.t.\ }\mcM(G(\bfz)+\bfe) = \bfy.
	\end{align*}
	Our problem is fundamentally different from the above problem. On the one hand, although the problem in \cite{dhar_modeling_2018} can also be treated as an outlier detection problem, the outlier occurs in {{signal $\bfx$ itself}}. However, in our problem, the {{outliers}} appear in the measuring process. On the other hand, {{our recovery performance guarantees are very different, and so are our analytical techniques.  
	}}
	}}

\section{Solving $\ell_1$ Minimization via Gradient Descent and Alternating Direction Method of Multipliers}\label{Sec:ADMMAlgorithm}

	In this section, we introduce both gradient descent algorithm and alternating direction method of multipliers (ADMM) algorithm to solve the outlier detection problem.

	We consider the $\ell_1$ minimization in Section \ref{Sec:ProblemStatement}, i.e.,
	\begin{align}\label{Defn:L1Minimization}
    \min_{\textbf{z}\in\mathbb{R}^k} \|{\cal M}G\left( {\bf{z}} \right)-\textbf{y}\|_1,
    \end{align}
    where $G(\cdot)$ is a well trained generator. 
{{
    Note that there can be different models for solving the outlier detection problem. We consider solving other problem models with different methods. For $\ell_1$ minimization, we consider both the case where we solve
\begin{align}\label{Defn:UnSquaredL1MinimizationViaADMM}
\min_{\z\in\mathbb{R}^k} \|\mathcal{M}G(\z) - \y\|_1
\end{align}
by the ADMM algorithm introduced, and the case where we solve
\begin{align}\label{Defn:SquaredL1MinimizationViaGD}
\min_{\z\in\mathbb{R}^k} \|\mathcal{M}G(\z) - \y\|_1^2
\end{align}
by the gradient descent (GD) algorithm \cite{Barzilai88}. 
For $\ell_2$ minimization, we {solve the following optimization problems:}
\begin{align}
\min_{\z\in\mathbb{R}^k} \|\mathcal{M}G(\z) - \y\|_2^2,
\end{align}
and
\begin{align}
\min_{\z\in\mathbb{R}^k} \|\mathcal{M}G(\z) - \y\|_2^2 + \lambda \|\z\|_2^2.
\end{align}
Both the above two $\ell_{2}$ norm minimizations are solved by gradient descent solver \cite{Barzilai88}.

}}

\subsection{Gradient Descent Algorithm}

    {{Theoretically speaking, due to the non-differentiability of the $\ell_1$ norm at $0$, the gradient descent method cannot be applied to solve the problem (\ref{Defn:L1Minimization}). Our numerical experiments also show that direct GD for $\ell_1$ norm does not yield good reconstruction results. However, once we turn to solve an equivalent problem, i.e.,
    \begin{align}\label{Defn:SquaredL1Minimization_State}
    \min_{\textbf{z}\in\mathbb{R}^k} \|{\cal M}G\left( {\bf{z}} \right)-\textbf{y}\|_1^2,
    \end{align}
    the gradient descent algorithm can solve (\ref{Defn:SquaredL1Minimization_State}) in practice though $\ell_1$ norm is not differentiable. Please see Section \ref{Sec:NumericalResults} for experimental results.}}  

\subsection{Iterative {{Linearized}} Alternating Direction Method of Multipliers Algorithm}

    Due to lack of theoretical guarantees, we propose an iterative linearized ADMM algorithm for solving (\ref{Defn:L1Minimization}) where the nonlinear mapping $G(\cdot)$ at each iteration is approximated via a linearization technique. Specifically, we introduce an auxiliary variable $\bfw$, and the above problem can be re-written as
\begin{align}\label{eq:min_l1_equivalent}
& \min_{\textbf{z}\in\mathbb{R}^k,\textbf{w}\in\mathbb{R}^{m}} \|\textbf{w}\|_1\nonumber\\
& {\rm s.t.\ }{\cal M}G(\textbf{z})-\textbf{w}=\textbf{y}.
\end{align}
{{
The Augmented Lagrangian of (\ref{eq:min_l1_equivalent}) is given as
\begin{align*}
L_{\rho}(\bfz,\bfw,\lambda) = \|w\|_{1} + \lambda^{T}(\mcM G(\bfz) - \bfw - \bfy) + \frac{\rho}{2} \| \mcM G(\bfz) - \bfw - \bfy\|_{2}^{2},
\end{align*}
or 
{\begin{align*}
L_\rho(\textbf{z},\textbf{w},{\bf \lambda})
& = \|\bfw\|_1 
+ \frac{\rho}{2}\left\|\mcM G(\bfz) - \bfw - \bfy + \frac{{\bf \lambda}}{\rho}\right\|^2_2 -\frac{\rho}{2} \|\lambda/\rho\|_{2}^{2},
\end{align*}}
where {{${\bf \lambda}\in\mathbb{R}^m$}} and $\rho\in\mathbb{R}$ are Lagrangian multipliers and penalty parameter respectively. The main idea of ADMM algorithm is to {update $\bfz$ and $\bfw$ alternatively}. Notice that for both $\bfw$ and ${\bf \lambda}$, they can be updated following the standard procedures. The nonlinearity and the lack of explicit form of the generator make it nontrivial to find the updating rule for $\bfz$. Here we consider a local linearization technique to solve the problem. 

Consider the $q$-th iteration for $\bfz$, we have 
\begin{align}
{{\bf z}^{q + 1}} 
&= \arg \mathop {\min }\limits_{\bf z} {L_\rho }\left( {{\bf z},{{\bf w}^q},{\lambda ^q}} \right) \nonumber\\
&= \arg \mathop {\min }\limits_{\bf z} F(\bfz),
\end{align}
where $F\left( {\bf z} \right)$ is defined as
$$
F(\bfz) = \left\| {{{\mathcal{M}}}G\left( {\bf z} \right) - {{\bf w}^q} - {\bf y} {{+}} \frac{{{\lambda ^q}}}{\rho }} \right\|_2^2.
$$
As we mentioned above, since $G({\bf z})$ is non-linear, we will use a first order approximation, i.e.,
\begin{align}
G\left( {\bf z} \right)\approx G({\bf z}^q)+\nabla G({\bf z}^q)^T({\bf z}-{\bf z}^q),
\end{align}
where $\nabla G({\bf z}^q)^T$ is the transpose of the gradient at ${\bf z}^q$. Then $F({\bf z})$ will be
\begin{align}
 F\left( {\bf z} \right)&=\left\| {{\cal M}\left( G({\bf z}^q)+\nabla G({\bf z}^q)^T ({\bf z}-{\bf z}^q) \right) - {{\bf w}^q} - {\bf y} {{+}} \frac{{{\lambda ^q}}}{\rho }} \right\|_2^2 \nonumber\\
&=\left\| {\cal M}\nabla G({\bf z}^q)^T{\bf z}+{\cal M}\left( G({\bf z}^q)-\nabla G({\bf z}^q)^T {\bf z}^q\right) - { {{\bf w}^q}} - {\bf y} {{+}} \frac{{{\lambda ^q}}}{\rho } \right\|_2^2.
\end{align}
The considered problem is a least square problem, and {the minimum is achieved at
\begin{align*}
\bfz^{q+1}
& = 
(\mcM\nabla G({\bf z}^q)^T )^{\dagger}
\left({\bf w}^q +{\bf y} - \frac{\lambda ^q}{\rho}-\left(\mcM G({\bf z}^q)-\mcM\nabla G({\bf z}^q)^T {\bf z}^q\right)\right),
\end{align*}
where $\dagger$ denotes the pseudo-inverse.} Notice that both  $\nabla G({\bf z}^q)$ and $G(\bfz^q)$ will be automatically computed by the Tensorflow \cite{abadi_tensorflow:_2016}. Similarly, we can update $\bfw$ and ${\bf \lambda}$ by
\begin{align}
{{\bf w}^{q + 1}} &= \arg \mathop {\min }\limits_{\bf w} \frac{\rho }{2}\left\| {{{\mathcal{M}}}G\left( {\bf z}\right) - {{\bf w}^q} - {\bf y} {{+}} \frac{{{\lambda ^q}}}{\rho }} \right\|_2^2 + {\left\| {\bf w}\right\|_1}\nonumber \\
&= {T_{\frac{1}{\rho }}}\left( {{\cal M}G\left( {{{\bf z}^{q + 1}}} \right) - {\bf y} {{+}} \frac{{{\lambda ^q}}}{\rho }} \right),
\end{align}
and 
\begin{align}
{\lambda ^{q + 1}} = {\lambda ^q} + \rho \left( {{\cal M}G\left( {{{\bf z}^{q + 1}}} \right) - {{\bf w}^{q + 1}} - {\bf y}} \right),
\end{align}
where ${T_{\frac{1}{\rho }}}$ is the element-wise soft-thresholding operator with parameter $\frac{1}{\rho}$ \cite{Boyd:2011}, i.e., 
\begin{align*}
[T_{\frac{1}{\rho}}(\bfx)]_i
= 
\begin{cases}
[\bfx]_i - \frac{1}{\rho}, [\bfx]_i> \frac{1}{\rho}, \\
0, |[\bfx]_i| \leq \frac{1}{\rho}, \\
[\bfx]_i + \frac{1}{\rho}, [\bfx]_i < - \frac{1}{\rho}.
\end{cases}
\end{align*}
}}

We continue ADMM algorithm to update $\bf z$ until optimization problem {{converges}} or stopping criteria are met. More discussion on stopping criteria and parameter tuning can be found in \cite{Boyd:2011}. {{Our algorithm is summarized in Algorithm \ref{Alg:IADMM}.}} The final value ${\bf z}^*$ can be used to generate the estimate $\hat {\bf x} = G({\bf z^*})$ for the true signal. We define the following metrics for evaluation: the measured error caused by imperfectness of CS measurement, i.e.,
\begin{align}\label{Defn:MeasureError}
{\varepsilon _m} \triangleq \left\| {\textbf{y} - {\cal M}G\left( {{\bf{\hat z}}} \right)} \right\|_1,
\end{align}
and reconstructed error via a mismatch of $G(\hat{\bf z})$ and $x$, i.e.,
\begin{align}\label{Defn:ReconstructionError}
{\varepsilon _r} \triangleq \left\| {{\bf{x}} - G\left( {{\bf{\hat z}}} \right)} \right\|_2^2.
\end{align}

\begin{algorithm}
\caption{{Linearized ADMM}}\label{Alg:IADMM}
\begin{algorithmic}

\STATE {\bf Input:} $\bfz^0, \bfw^0, {\bf \lambda}^0, \mcM$

\STATE {\bf Parameters:} $\rho$, $MaxIte$

\STATE $q \gets 0$

\WHILE{$q\leq MaxIte$}

\STATE $A^{q+1} \gets \mcM \nabla G(\bfz^{q})^T$

\STATE $\bfz^{q+1} \gets {(A^{q+1})^{\dagger}}(\bfw^q + \bfy - {\bf\lambda}^q/\rho - (\mcM G(\bfz^q) - A^{q+1} \bfz^q))$

\STATE $\bfw^{q+1} \gets T_{\frac{1}{\rho}}\left( \mcM G(\bfz^{q+1}) - \bfy + \frac{{\bf \lambda}^q}{\rho} \right)$

\STATE ${\bf\lambda}^{q+1} \gets {\bf\lambda}^q + \rho (\mcM G(\bfz^{q+1}) - \bfw^{q+1} - \bfy)$

\ENDWHILE

\STATE {\bf Output:} $z^{MaxIte+1}$

\end{algorithmic}
\end{algorithm}

\section{Performance Analysis}\label{Sec:PerformanceAnalysis}

In this section, we present theoretical analysis {{of}} the performance of signal reconstruction method based on the generative model. We first establish the necessary and sufficient conditions for recovery in {{Theorems}} \ref{Thm:L0MinimizationRecoveryCondition} and \ref{Thm:L1MinimizationRecoveryGuarantee}. Note that these two theorems appeared in \cite{xu_sparse_2013}, and we present them and their proofs here for the sake of completeness. We then show that both the linear and nonlinear neural network with {{an}} arbitrary {{number of}} layers can satisfy the recovery condition, thus {ensuring} successful reconstruction of signals.

\subsection{Necessary and Sufficient Recovery Conditions}

{Suppose} the ground truth signal $\textbf{x}_0\in\mathbb{R}^n$, the measurement matrix ${\cal M}\in\mathbb{R}^{m\times n} ({m < n})$, and the sparse outlier vector $\textbf{e}\in\mathbb{R}^{{m}}$ such that $\|\textbf{e}\|_0\leq {l < m}$, then the problem can be stated as
\begin{align}\label{Defn:L0Minimization}
\min_{z\in\mathbb{R}^k} \|{\cal M}G(\textbf{z}) -\textbf{ y}\|_0
\end{align}
where $\textbf{y} = {\cal M}\textbf{x}_0+\textbf{e}$ is a measurement vector and $G(\cdot):\mathbb{R}^k\rightarrow \mathbb{R}^n$ is a generative model. Let $\textbf{z}_0\in\mathbb{R}^k$ be a vector such that $G(\textbf{z}_0)=\textbf{x}_0$, then we have following conditions under which the $\textbf{z}_0$ can be recovered exactly without any sparsity assumption on neither $\textbf{z}_0$ nor $\textbf{x}_0$.

\begin{theorem}\label{Thm:L0MinimizationRecoveryCondition}
	Let $G(\cdot), \textbf{x}_0, \textbf{z}_0, \textbf{e}$ and $\textbf{y}$ be as above. The vector $\textbf{z}_0$ can be recovered exactly from (\ref{Defn:L0Minimization}) {for any $\textbf{e}$ with $\|\textbf{e}\|_0\leq l$} if and only if $\|{\cal M}G(\textbf{z})-{\cal M}G(\textbf{z}_0)\|_0 \geq 2l+1$ holds for any $\textbf{z}\neq \textbf{z}_0$.
\end{theorem}

{\emph Proof:}
    We first show the sufficiency. Assume $\|{\cal M}G(\textbf{z})-{\cal M}G(\textbf{z}_0)\|_0 \geq 2l+1$ holds for any $\textbf{z}\neq \textbf{z}_0$, since the triangle inequality gives
	$$
	\|{\cal M}G(\textbf{z}) - {\cal M}G(\textbf{z}_0)\|_0
	\leq \|{\cal M}G(\textbf{z}_0)-\textbf{y}\|_0 + \|{\cal M}G(\textbf{z})-\textbf{y}\|_0,
	$$
	then
	\begin{align*}
	\|{\cal M}G(\textbf{z}) - \textbf{y}\|_0
	& \geq \|{\cal M}G(\textbf{z}) - {\cal M}G(\textbf{z}_0)\|_0 - \|{\cal M}G(\textbf{z}_0)-\textbf{y}\|_0 \\
&	= (2l+1)-l \\
&	\geq l+1 \\
&	> \|{\cal M}G(\textbf{z}_0) - \textbf{y}\|_0,
	\end{align*}
	and this means $\textbf{z}_0$ is {{a}} unique optimal solution to (\ref{Defn:L0Minimization}).
	
	Next, we show the necessity by contradiction. Assume $\|{\cal M}G(\textbf{z}) - {\cal M}G(\textbf{z}_0)\|_0\leq 2l$ holds for certain $\textbf{z}\neq \bfz_0\in\mathbb{R}^k$, and this means ${\cal M}G(\textbf{z})$ and ${\cal M}G(\textbf{z}_0)$ differ from each other over at most $2l$ entries. Denote by $\mathcal{I}$ the index set where ${\cal M}G(\textbf{z}) - {\cal M}G(\textbf{z}_0)$ has nonzero entries, then $|\mathcal{I}|\leq 2l$. We can choose outlier vector $\textbf{e}$ such that $\textbf{e}_i=({\cal M}G(\textbf{z}) - {\cal M}G(\textbf{z}_0))_i$ for all $i\in\mathcal{I}'\subset \mathcal{I}$ with $|\mathcal{I}'|= l$, and $\textbf{e}_i=0$ otherwise. Then
	\begin{align}
	\|{\cal M}G(\textbf{z}) - \textbf{y}\|_0
	= \|{\cal M}G(\textbf{z}) - {\cal M}G(\textbf{z}_0) - \bfe\|_0
	\leq 2l - l
	=l = \|{\cal M}G(\textbf{z}_0) - \textbf{y}\|_0,
	\end{align}
	which means we find an optimal solution $\bfz$ to (\ref{Defn:L0Minimization}) which is not $\bfz_0$. This contradicts the assumption.

$\hfill\square$

Theorem \ref{Thm:L0MinimizationRecoveryCondition} gives a necessary and sufficient condition for successful recovery via $\ell_0$ minimization, and we also present {{an}} equivalent condition as stated in Theorem \ref{Thm:L1MinimizationRecoveryGuarantee} for {{successful recovery via $\ell_1$ minimization.}}

\begin{theorem}\label{Thm:L1MinimizationRecoveryGuarantee}
	Let $G(\cdot), \bfx_0, \bfz_0, \bfe, {\cal M} $, and $\bfy$ be the same as in Theorem \ref{Thm:L0MinimizationRecoveryCondition}. The low dimensional {{vector}} $\bfz_0$ can be recovered correctly from any $\e$ with $\|\e\|_0\leq l$ from
	\begin{equation}\label{eqn:norm1}
	\min_{\z} \quad \|\y-{\cal M}G(\z)\|_1,
	\end{equation}
	if and only if for any $\bfz \neq \bfz_0$,
	$$
	\|({\cal M}G(\bfz_0)-{\cal M}G(\bfz))_{K}\|_{1} < \|({\cal M}G(\bfz_0)-{\cal M}G(\bfz))_{\overline{K}}\|_{1},
	$$
	where $K$ is the support of the error vector $\e$.
	
\end{theorem}

{\emph Proof:} We first show the sufficiency, i.e., if the $\|({\cal M}G(\bfz_0)-{\cal M}G(\bfz))_{K}\|_{1} < \|({\cal M}G(\bfz_0)-{\cal M}G(\bfz))_{\overline{K}}\|_{1}$ holds for any $\bfz \neq \bfz_0$ where $K$ is the support of $\e$, then we can correctly recover $\bfz_0$ from (\ref{eqn:norm1}). For a feasible solution $\bfz$ to (\ref{eqn:norm1}) which differs from $\bfz_0$, we have
	\begin{align*}
		\|\y- {\cal M}G(\z)\|_{1}
		&= \|({\cal M}G(\z_0)+\e)- {\cal M}G(\z)\|_{1}\\
		&= \|\e_{K}- ({\cal M}G(\z)-{\cal M}G(\z_0))_{K}\|_{1}
		+
		\|({\cal M}G(\z)-{\cal M}G(\z_0))_{\overline{K}}\|_{1}\\
		&\geq \|\e_{K}\|_{1}- \|({\cal M}G(\z)-{\cal M}G(\z_0))_{K}\|_{1}
		+
		\|({\cal M}G(\z)-{\cal M}G(\z_0))_{\overline{K}}\|_{1}\\
		&> \|\e_{K}\|_{1}=\|\y-{\cal M}G(\z_0)\|_1,
	\end{align*}
	which means that $\z_0$ is the unique optimal solution to (\ref{eqn:norm1}) and can be exactly recovered by solving (\ref{eqn:norm1}).
	
	We now show the necessity by contradiction. Suppose that there exists an $\z \neq \z_0$ such that $\|({\cal M}G(\z_0)-{\cal M}G(\z))_{K}\|_{1} \geq \|({\cal M}G(\z_0)-{\cal M}G(\z))_{\overline{K}}\|_{1}$ with $K$ being the support of $\e$. Then we can pick $\e_i$ to be $({\cal M}G(\z)-{\cal M}G(\z_0))_i$ if $i\in K$ and $0$ if $i\in\overline{K}$. Then
	\begin{align*}
		\|\y- {\cal M}G(\z)\|_{1}
		&= \|{\cal M}G(\z_0)- {\cal M}G(\z)+\e\|_{1}\\
		&= \|({\cal M}G(\z)-{\cal M}G(\z_0))_{\overline{K}}\|_{1}\\
		&\leq \|({\cal M}G(\z_)-{\cal M}G(\z))_{K}\|_{1}\\
		& =\|\e\|_1
		=\|y-{\cal M}G(\z_0)\|_1,
	\end{align*}
	which means that $\z_0$ cannot be the unique optimal solution to (\ref{eqn:norm1}). Thus
	$$
	\|({\cal M}G(\z)-{\cal M}G(\z_0))_{\overline{K}}\|_{1}
		> \|({\cal M}G(\z_)-{\cal M}G(\z))_{K}\|_{1}
	$$
	must holds for any $\bfz\neq \bfz_0$.

$\hfill\square$

{{ \subsection{Full Rankness of Random Matrices} }}


Before we proceed to the main results, we give some technical lemmas which will be used to prove our main theorems.

\begin{lemma}\label{Lem:Schwartz-ZippelLemma}
(\cite{zippel_effective_2012}) Let $P\in A[X_1,\cdots,X_v]$ be a polynomial of total degree $D$ over an integral domain $A$. Let $\mathscr{J}$ be a subset of $A$ of cardinality $B$. Then
$${
  \mathbb{P}(P(x_1,\cdots,x_v)=0|x_i\in\mathscr{J}) \leq \frac{D}{B}.
}$$
\end{lemma}

Lemma \ref{Lem:Schwartz-ZippelLemma} gives an upper bound of the probability for a polynomial to be zero when its variables are randomly taken from a set. Note that in \cite{khajehnejad_sparse_2011}, the authors applied the Lemma \ref{Lem:Schwartz-ZippelLemma} to study the adjacency matrix from an expander graph. Since the determinant of a square random matrix is a polynomial, Lemma \ref{Lem:Schwartz-ZippelLemma} can be applied to determine whether a square random matrix is rank deficient. An immediate result will be Lemma \ref{Lem:FullRankOfGaussianRandomMatrix}.

\begin{lemma}\label{Lem:FullRankOfGaussianRandomMatrix}

A random matrix $M\in\mathbb{R}^{m\times n}$ with independent Gaussian entries will have full rank with probability $1$, i.e., ${\rm rank}(M)=\min(m,n)$.

\end{lemma}

{\emph Proof:} For simplicity of presentation, we consider the square matrix case, i.e., a random matrix $M\in\mathbb{R}^{n\times n}$ whose entries are drawn i.i.d. randomly from $\mathbb{R}$ according to standard Gaussian distribution $\mathcal{N}(0,1)$.

Note that
$$
{\rm det}(M) = \sum_{\sigma} \left({\rm sign}(\sigma) \prod_{i=1}^n M_{i\sigma(i)}\right)
$$
where $\sigma$ is a permutation of $\{1,2,\cdots,n\}$, the summation is over all $n!$ permutations, and ${\rm sign}(\sigma)$ is either $+1$ or $-1$. Since ${\rm det}(M)$ is a polynomial of degree $n$ with respect to $M_{ij}\in\mathbb{R}$, $i,j=1,\cdots,n$ and the $\mathbb{R}$ {{has infinitely many elements}}, then according to Lemma \ref{Lem:Schwartz-ZippelLemma}, we have
\begin{align*}
\mathbb{P}({\rm det}(M)=0) \leq 0.
\end{align*}
Thus
\begin{align*}
\mathbb{P}(M\text{ has full rank})
= \mathbb{P}({\rm det}(M)\neq 0)
\geq 1 - \mathbb{P}({\rm det}(M)=0),
\end{align*}
and this means the Gaussian random matrix has full rank with probability $1$.

$\hfill\square$

{\bf Remarks:} (1) We can easily extend the arguments to random matrix with arbitrary shape, i.e., $M\in\mathbb{R}^{m\times n}$. {{Considering}} arbitrary square sub-matrix with size $\min(m,n)\times \min(m,n)$ from $M$, following the same arguments will give that with probability 1, the matrix $M$ has full rank with ${\rm rank}(M)=\min(m,n)$; (2) The arguments in Lemma \ref{Lem:FullRankOfGaussianRandomMatrix} can also be {{extended}} to random matrix with other distributions.

\subsection{Generative {{Models}} via Linear Neural {{Networks}}}

We first consider the case where the activation function is an element-wise identity operator, i.e., $a(\bfx)=\bfx$, and this will give us a simplified input and output relation as follows
\begin{align}\label{Eq:LinearNN}
G(\bfz) = W^{(d)}W^{(d-1)}\cdots W^{(1)}\bfz.
\end{align}
Define $W = W^{(d)}W^{(d-1)}\cdots W^{(1)}$, we get a simplified relation $G(\bfz)=W\bfz$. We will show that with high probability: for any $\bfz\neq \bfz_0$, the
$$
\|G(\bfz) - G(\bfz_0)\|_0
= \|W(\bfz - \bfz_0)\|_0
$$
will have at least $2l+1$ nonzero entries; or when we treat the $\z-\z_0$ as a whole and still use the letter $\z$ to denote $\z-\z_0$, then for any $\bfz\neq 0$, the
$$
\bfv := W\bfz
$$
will have at least $2l+1$ nonzero entries.

Note that in the above derivation, the bias term is incorporated in the weight matrix. For example, in the first layer, we know
$$
W^{(1)}\bfz + \bfb^{(1)}
= [W^{(1)}\ \bfb^{(1)}]
\left[\begin{matrix}
\bfz \\ 1
\end{matrix}\right].
$$
Actually, when the $W^{(i)}$ does not incorporate the bias term ${\bf b}^{(i)}$, we have a generator as follows
\begin{align*}
G(z)
& = W^{(d)}\left(W^{(d-1)}\left(
                                \cdots \left(W^{(2)}\left(W^{(1)}{\bf z} + {\bf b}^{(1)}\right) + {\bf b}^{(2)}
                                            \right)
                                \right) +\cdots + {\bf b}^{(d-1)}
                \right) + {\bf b}^{(d)}  \\
& = W^{(d)}\cdots W^{(1)}\z
+ W^{(d)}\cdots W^{(2)} {\bf b}^{(1)}
+ W^{(d)}\cdots W^{(3)} {\bf b}^{(2)} \\
& \ \ \ \
+ \cdots
+ W^{(d)}{\bf b}^{(d-1)} + {\bf b}^{(d)}.
\end{align*}
We can define $W = W^{(d)}\cdots W^{(1)}$. Then in this case, we need to show that with high probability: for any $\z\neq \z_0$, the
\begin{align*}
\|G(\z) - G(\z_0)\|_0
& = \|W(\z - \z_0)\|_0
\end{align*}
{has at least $2l+1$ nonzero entries because all the other terms are cancelled.} 

{{
\begin{lemma}\label{Lem:FullRankAndZeroEntries}
Let matrix $W\in\mbR^{n\times k}$. Then for an integer $l \geq 0$ , if every $r:= n-(2l+1)\geq k$ rows of $W$ has full rank $k$, then $\bfv =W \bfz$ can have at most $n-(2l+1)$ zero entries {{for every}} $\bfz\neq 0\in\mathbb{R}^k$.
\end{lemma}
}}

{\emph Proof:} Suppose $\bfv = W\bfz$ has $n-(2l+1)+1$ zeros, {{we}} denote by $S_\bfv$ the set of indices where the corresponding entries of $\bfv$ are nonzero, and by $\bar{S}_\bfv$ the set of indices where the corresponding entries of $\bfv$ are zero, then $|\bar{S}_\bfv|=n-(2l+1)+1$. Then for the homogeneous linear equations
$${
  \bm{0} = \bfv_{\bar{S}_\bfv} = W_{\bar{S}_\bfv} \bfz,
}$$
where $W_{\bar{S}_\bfv}$ consists of $|\bar{S}_\bfv| = n-(2l+1) + 1$ rows from $W$ with indices in $\bar{S}_\bfv$, the matrix $W_{\bar{S}_\bfv}$ must have a rank less than $k$ so that there exists {{such}} a nonzero solution $\bfz$. Then there {{exists}} a sub-matrix consisting of $n-(2l+1)$ rows which has rank less than $k$, and this contradicts the statement.

$\hfill\square$

The Lemma \ref{Lem:FullRankAndZeroEntries} actually gives a sufficient condition for $\bfv = W\bfz$ to have at least $2l+1$ nonzero entries, i.e., every $n-(2l+1)$ rows of $W$ has full rank $k$.

\begin{lemma}\label{Lem:LinearNNArbitraryLayers}

Let the composite weight matrix in multiple-layer neural network with identity activation function be
$${
  W = W^{(d)}W^{(d-1)}\cdots W^{(1)},
}$$
where
$$
W^{(1)} = [w^{(1)}_1,w^{(1)}_2,\cdots,w^{(1)}_{n_1}]^T\in\mathbb{R}^{n_1 \times k}, w_j^{(1)}\in\mbR^{k}, j=1,\cdots,n_1, 
$$
$$
W^{(i)}
= [w^{(i)}_1,w^{(i)}_2,\cdots,w^{(i)}_{n_i}]^T\in\mathbb{R}^{n_i \times n_{i-1}}, w_j^{(i)}\in\mbR^{n_{i-1}}, 1< i < d,{j=1\cdots,n_i,}
$$
and
$$
W^{(d)} = [w^{(d)}_1,w^{(d)}_2,\cdots,w^{(d)}_{n}]^T\in\mathbb{R}^{n \times n_{d-1}}, w_j^{(d)}\in\mbR^{n_{d-1}}, j=1,\cdots,n,
$$
with $n_i\geq k, i=1,\cdots,d-1$  and {{$n> k$}}. When $d=1$, we let $W = W^{(1)}\in\mathbb{R}^{n\times k}$. Let each entry in each weight matrix $W^{(i)}$ be drawn independently randomly according to {{the standard Gaussian distribution, $\mcN(0,1)$}}. Then for a linear neural network {{of}} two or more layers {{with}} composite weight matrix $W$ defined above, with probability 1, every $r=n-(2l+1)\geq k$ rows of $W$ will have full rank $k$.

\end{lemma}

{\emph Proof:} We will show Lemma \ref{Lem:LinearNNArbitraryLayers} by induction.

{\bf 2-layer case:} When there are only two layers in the neural network, we have
$$
W = W^{(2)}W^{(1)}
$$
where $W^{(1)}\in\mathbb{R}^{n_1\times k}$ ($n_1\geq k$) and $W^{(2)}\in\mathbb{R}^{n\times n_1}$ ($n\geq n_1)$. From Lemma \ref{Lem:FullRankOfGaussianRandomMatrix}, with probability $1$, the matrix $W^{(1)}$ has full rank and a singular value decomposition
$$
W^{(1)} = U^{(1)} \Sigma^{(1)} (V^{(1)})^*,
$$
where $\Sigma^{(1)}\in\mathbb{R}^{k\times k}$ and $V^{(1)}\in\mathbb{R}^{k\times k}$ have rank $k$, and $U^{(1)}\in\mathbb{R}^{n_1\times k}$.

For the matrix $W^{(2)}W^{(1)}$, we take arbitrary $r=n-(2l+1)\geq k$ rows of $W^{(2)}$ to form a new matrix $M^{(2)}$, and we have
$$
l \leq \frac{n-1-k}{2},
$$
and
\begin{align*}
M^{(2)}W^{(1)}
= M^{(2)}U^{(1)}\Sigma^{(1)} (V^{(1)})^*.
\end{align*}
Since $\Sigma^{(1)}$ and $V^{(1)}$ have full rank $k$, then
$$
{\rm rank}(M^{(2)} W^{(1)})
= {\rm rank}(M^{(2)}U^{(1)}).
$$

{{If we}} fix the matrix $W^{(1)}$, then $U^{(1)}$ will also be fixed. {The matrix $M^{(2)}U^{(1)}$ has full rank $k$ with probability 1, so is $M^{(2)}W^{(1)}$.} {{Notice that we have totally ${n\choose{n-(2l+1)}}$ choices for forming $M^{(2)}$, thus according to the union bound, the probability for existence of a matrix $M^{(2)}$ which makes $M^{(2)}W^{(1)}$ rank deficient satisfies
$$
\mbP(\text{there exist a matrix } M^{(2)} \text{ such that } M^{(2)}W^{(1)} \text{ is rank deficient}) \leq {n\choose{n-(2l+1)}} \times 0 =0.
$$
Thus, the probability for all such matrices $M^{(2)}$ to have full rank is $1$.}}

{\bf $(d-1)$-layer case:} Assume for the case with $d-1$ layers, arbitrary $n-(2l+1)$ rows of the weight matrix
$$
W = W^{(d-1)}W^{(d-2)}\cdots W^{(1)}
$$
will form a matrix with full rank $k$, where $W^{(d-1)}\in\mathbb{R}^{n\times n_{d-2}}$ and $W^{(d-2)}\in\mathbb{R}^{n_{d-2}\times n_{d-3}}$. Then for $W^{(d-2)}\cdots W^{(1)}$, it will have rank $k$ and singular value decomposition
$$
W^{(d-2)}\cdots W^{(1)} = U \Sigma V^*,
$$
where $U\in\mathbb{R}^{n_{d-2}\times k}$, $\Sigma\in\mathbb{R}^{k\times k}$ and $V\in\mathbb{R}^{k\times k}$.

{\bf $d$-layer case:} Now we consider the case with $d$ layers. The corresponding composite weight matrix becomes
\begin{align*}
W = W^{(d)}W^{(d-1)}W^{(d-2)}\cdots W^{(1)}
\end{align*}
where $W^{(d)}\in\mathbb{R}^{n\times n_{d-1}}$, $W^{(d-1)}\in\mathbb{R}^{n_{d-1}\times n_{d-2}}$ and $W^{(d-2)}\in\mathbb{R}^{n_{d-2}\times n_{d-3}}$.

Since
$$
W^{(d-2)}\cdots W^{(1)} = U \Sigma V^*,
$$
then
\begin{align*}
W = W^{(d)}W^{(d-1)}U \Sigma V^*.
\end{align*}
Note that
$$
{\rm rank}(W^{(d-1)} U\Sigma V^*) = {\rm rank}(W^{(d-1)}U),
$$
{$W^{(d-1)}U$ is a matrix of full rank $k$ with probability 1} and has a singular value decomposition as
$$
W^{(d-1)}U = \tilde{U}\tilde{\Sigma}\tilde{V}^*,
$$
where $\tilde{U}\in\mathbb{R}^{n_{d-1}\times k}$, $\tilde{\Sigma}\in\mathbb{R}^{k\times k}$ and $\tilde{V}\in\mathbb{R}^{k\times k}$. Follow the similar arguments above, {{every}} $n-(2l+1)$ rows of $W^{(d)}W^{(d-1)}W^{(d-2)}\cdots W^{(1)}$ will have full rank $k$.

$\hfill\square$

{\bf Remarks:} Actually the statement also holds when $d=1$. When the NN has only 1 layer, we have $W=W^{(1)}$. Then matrix $M^{(1)}$ consisting of arbitrary $r$ rows in $W$ will also be a random matrix with each entry drawn i.i.d. randomly according to Gaussian distribution, thus with probability 1, the matrix $M^{(1)}$ will have full rank $k$.

Thus, we can conclude that for every $\bfz\neq 0\in\mathbb{R}^k$, if the weight matrices are defined {{as}} above, then the $y=Wx\in\mathbb{R}^n$ will have at least $2l+1$ nonzero elements, and the successful recovery is guaranteed by the following theorem.

\begin{theorem}\label{Thm:LinearNNRecoveryGuarantee}

{{
Let the generator $G(\cdot)$ be implemented by a multiple-layer neural network with identity activation function. The weight matrix in each layer is defined as
$$
W^{(1)} = [w^{(1)}_1,w^{(1)}_2,\cdots,w^{(1)}_{n_1}]^T\in\mathbb{R}^{n_1 \times k}, w_j^{(1)}\in\mbR^{k}, j=1,\cdots,n_1
$$
$$
W^{(i)}
= [w^{(i)}_1,w^{(i)}_2,\cdots,w^{(i)}_{n_i}]^T\in\mathbb{R}^{n_i \times n_{i-1}}, w_j^{(i)}\in\mbR^{n_{i-1}}, 1< i < d, {j=1\cdots,n_i,} 
$$
and
$$
W^{(d)} = [w^{(d)}_1,w^{(d)}_2,\cdots,w^{(d)}_{n}]^T\in\mathbb{R}^{n \times n_{d-1}}, w_j^{(d)}\in\mbR^{n_{d-1}}, j=1,\cdots,n,
$$
with $n_i\geq k, i=1,\cdots,d-1$  and {{$n> k$}}. Each entry in each weight matrix $W^{(i)}$ is independently randomly drawn from standard Gaussian distribution $\mcN(0,1)$. Let the measurement matrix $\mathcal{M}\in\mathbb{R}^{m\times n}$ be a random matrix with each entry iid drawn according to standard Gaussian distribution $\mcN(0,1)$. Let the true signal $\bfx_0$ and its corresponding low dimensional signal $\bfz_0$, the outlier signal $\bfe$ and $\y$ be the same as in Theorem \ref{Thm:L0MinimizationRecoveryCondition} and {{satisfy
$$
\|\bfe\|_{0} \leq \frac{n-1-k}{2}.
$$ 
}}
Then with probability $1$: for a well-trained generator $G(\cdot)$, every $\bfz_0$ 
can be recovered from
$$
\min_{\bfz \in\mathbb{R}^k} \|\mathcal{M}G(\bfz) - y\|_0.
$$
}}

\end{theorem}

{\emph Proof:} Notice that in this case, we have
$$
\mathcal{M}G(\bfz) = \mathcal{M}W^{(d)}W^{(d-1)}\cdots W^{(1)}\bfz.
$$
Since $\mathcal{M}^{m\times n}$ is also a random matrix with i.i.d. Gaussian entries $l\leq \frac{n-1-k}{2}$, we can actually treat $\mathcal{M}G(\bfz)$ as a $(d+1)$-layer neural network implementation of the generator acting on $\bfz$. Then {{according}} to Lemma \ref{Lem:LinearNNArbitraryLayers}, every ${m}-(2l+1)$ rows of $\mathcal{M}W^{(d)}W^{(d-1)}\cdots W^{(1)}$ will have full rank $k$. The Lemma \ref{Lem:FullRankAndZeroEntries} implies that the $\mathcal{M}W^{(d)}W^{(d-1)}\cdots W^{(1)}(\z-\z_0)$ can have at most ${m}-(2l+1)$ zero entries for every $\z\neq \z_0$. Finally, from Theorem \ref{Thm:L0MinimizationRecoveryCondition}, the $\bfz_0$ can be recovered with probability 1.

$\hfill\square$

\subsection{Generative {{Models}} {{via}} Nonlinear Neural {{Network}}}

In previous sections, we have show that when the activation function $a(\cdot)$ is the identity function, we can show that with high probability, the generative model implemented via neural network can successfully recover the true signal $\bfz_0$. In this section, we will extend our analysis to the neural network with nonlinear activation functions.

{{
	Notice that in the proof of neural {{networks}} with identity activation, the key is that we can write
	$G(\bfz) - G(\bfz_0)$ as $W(\bfz-\bfz_0)$, which allows us to characterize the recovery conditions by the properties of linear equation systems. However, in neural {{network}} with nonlinear activation functions, we do not have such linear properties. For example, in the case {{of}} leaky ReLU activation function, i.e.,
$$
[a(\bfx)]_i = 
\begin{cases}
\bfx_i, \bfx_i \geq 0,\\
h \bfx_i, \bfx_i<0,
\end{cases}
$$
where $h\in(0,1)$, for an $\bfz\neq \bfz_0$, the sign patterns of $\bfz$ and $\bfz_0$ can be different, and this results in that the rows of the weight matrix in each layer will be scaled by different values. Thus we cannot directly transform $G(\bfz) - G(\bfz_0)$ into $W(\bfz - \bfz_0)$. Fortunately, we can achieve the transformation via Lemma \ref{Lem:PatternIrrelevanceInLeakyReLU} for the leaky ReLU activation function.

\begin{lemma}\label{Lem:PatternIrrelevanceInLeakyReLU}
 Define the leaky ReLU activation function as
$$
a(x) =
\begin{cases}
x, x\geq 0, \\
hx, x<0,
\end{cases}
$$
where $x\in\mbR$ and $h\in(0,1)$, then for arbitrary $x\in\mbR$ and $y\in\mbR$, the $a(x)-a(y)=\beta(x-y)$ holds for some {{$\beta\in[h,1]$}} regardless of the sign patterns of $x$ and $y$.
\end{lemma}

{\emph Proof:} When $x$ and $y$ have the same sign pattern, i.e., $x\geq 0$ and $y\geq 0$ holds simultaneously, or $x<0$ and $y<0$ holds simultaneously, then
$$
a(x) - a(y) = x - y,
$$
with $\beta=1$, or
$$
a(x) - a(y) = hx-hy = h(x-y),
$$
with $\beta=h\in(0,1)$.

When $x$ and $y$ have different sign patterns, we have the following two cases: (1) $x\geq 0$ and $y<0$ holds simultaneously; (2) $x<0$ and $y\geq 0$ holds simultaneously. For the first case, we have
\begin{align*}
a(x) - a(y)
 = x - hy
 = \beta (x-y),
\end{align*}
{{which gives
$$
\beta = \frac{x-hy}{x-y}
> \frac{hx - hy}{x-y}
 = h,
$$
and 
$$
\beta = \frac{x-hy}{x-y}
 < \frac{x-y}{x-y} = 1,
$$
where we use the facts that  $x\geq 0$, $y<0$ and $h\in(0,1)$.

Thus $\beta\in(h,1)$, and similarly for $x<0$ and $y\geq 0$. Combine the above cases together, we have $\beta\in[h,1]$.}}

$\hfill\square$

Since in the neural network, the leaky ReLU acts on the weighted sum $W\bfz$ {{element-wise}}, we can easily extend Lemma \ref{Lem:PatternIrrelevanceInLeakyReLU} to  the high dimensional case. For example, consider the first layer with weight $W^{(1)}\in\mbR^{{n_1}\times k}$, bias $\bfb^{(1)}\in\mbR^{{n_1}}$, and leaky ReLU activation function, {{we have
\begin{align*}
& a^{(1)}(W^{(1)}\bfz + \bfb^{(1)}) - a^{(1)}(W^{(1)}\bfz_0 + \bfb^{(1)}) \\
& =
\left[\begin{matrix}
\beta^{(1)}_1 ([W^{(1)}\bfz + \bfb^{(1)}]_1 - [W^{(1)}\bfz_0 + \bfb^{(1)}]_1) \\
\beta^{(1)}_2 ([W^{(1)}\bfz + \bfb^{(1)}]_2 - [W^{(1)}\bfz_0 + \bfb^{(1)}]_2) \\
\vdots\\
\beta^{(1)}_{{n_1}} ([W^{(1)}\bfz + \bfb^{(1)}]_{{n_1}} - [W^{(1)}\bfz_0 + \bfb^{(1)}]_{{n_1}})
\end{matrix}\right] \\
& =
\left[\begin{matrix}
\beta^{(1)}_1 & 0 &\cdots &0\\
0 & \beta^{(1)}_2 &\cdots &0\\
\vdots &\vdots &\ddots &0\\
0 & 0 & \cdots &\beta^{(1)}_{{n_1}}
\end{matrix}\right]
((W^{(1)}\bfz + \bfb^{(1)}) - (W^{(1)}\bfz_0 + \bfb^{(1)})) \\
& = \Gamma^{(1)} W^{(1)}(\bfz-\bfz_0),
\end{align*}
where $\beta^{(1)}_i\in[h,1]$ for every $i=1,\cdots,{n_1}$, and
$$
\Gamma^{(1)} =
\left[\begin{matrix}
\beta^{(1)}_1 & 0 &\cdots &0\\
0 & \beta^{(1)}_2 &\cdots &0\\
\vdots &\vdots &\ddots &0\\
0 & 0 & \cdots &\beta^{(1)}_{{n_1}}
\end{matrix}\right].
$$}}
Since $\Gamma^{(1)}$ has full rank, it does not affect the rank of $W^{(1)}$, and we can treat their product as a {{full rank}} matrix $P^{(1)} = \Gamma^{(1)}W^{(1)}$. In this way, we can apply the techniques in the linear neural networks to the one with leaky ReLU activation function.

{\bf Remarks:} Note that even though the Lemma \ref{Lem:PatternIrrelevanceInLeakyReLU} applies to ReLU activation function with $\beta\in[0,1]$, the $\Gamma$ matrix we construct can be rank deficient, which can cause the results in the linear neural network to fail; 

\begin{lemma}\label{Lem:LeakyReLUNNArbitraryLayers}

Let the generative model $G(\cdot):\mathbb{R}^k\to\mathbb{R}^n$ be implemented by a $d$-layer neural network, i.e.,
{{
$$
G(\bfz) = a^{(d)}\left(
      W^{(d)}a^{(d-1)}\left(
      W^{(d-1)} \cdots a^{(1)}\left(
                         W^{(1)}\bfz + \bfb^{(1)}
                        \right) \cdots + \bfb^{(d-1)}
             \right) + \bfb^{(d)}
\right).
$$}}
Let the weight matrix in each layer be defined as
$$
W^{(1)} = [w^{(1)}_1,w^{(1)}_2,\cdots,w^{(1)}_{n_1}]^T\in\mathbb{R}^{n_1 \times k}, w^{(1)}_j\in\mbR^k, j=1,\cdots,n_1,  
$$
$$
W^{(i)}
= [w^{(i)}_1,w^{(i)}_2,\cdots,w^{(i)}_{n_i}]^T\in\mathbb{R}^{n_i \times n_{i-1}}, 
w^{(i)}_j\in\mbR^{n_{i-1}}, 1< i < d, j=1,\cdots,n_i
$$
and
$$
W^{(d)} = [w^{(d)}_1,w^{(d)}_2,\cdots,w^{(d)}_{n}]^T\in\mathbb{R}^{n \times n_{d-1}},
w^{(d)}_j, j=1,\cdots,n,
$$
with 
{{
$$
n_i\geq k, i=1,\cdots,d-1,
$$
and 
$$
k\leq n-(2l+1).
$$ 
}}
Let each entry in each weight matrix $W^{(i)}$ be drawn iid randomly according to the standard Gaussian distribution $\mcN(0,1)$. Let the element-wise activation function $a^{(i)}$ in the $i$-th layer be a leaky ReLU, i.e.,
\begin{align*}
[a^{(i)}(x)]_j =
\begin{cases}
x_j, {\rm\ if\ }x_j \geq 0,\\
h^{(i)} x_j, {\rm\ if\ }x_j< 0,
\end{cases}
, i=1,\cdots,d,
\end{align*}
where $h^{(i)}\in(0,1)$.


Then {{with high probability, the following holds:}} for {{every}} $\bfz\neq \bfz_0$,
$$
\|G(\bfz_0) - G(\bfz)\|_0 \geq 2l+1,
$$
where $G(\cdot)$ is the generative model defined above.

\end{lemma}

{\emph Proof:} Apply the Lemma \ref{Lem:PatternIrrelevanceInLeakyReLU}, we have 
\begin{align*}
G(\bfz)
& = a^{(d)}\left(
      W^{(d)}a^{(d-1)}\left(
      W^{(d-1)} \cdots a^{(1)}\left(
                         W^{(1)}\bfz
                        \right)
             \right)
\right) \\
& =
\left[\begin{matrix}
h_1^{(d)} &0 &\cdots &0\\
0 & h_2^{(d)}  &\cdots &0\\
\vdots &\vdots &\ddots &\vdots\\
0 &0 &\cdots &h_{n}^{(d)}
\end{matrix}\right]
W^{(d)}
\left[\begin{matrix}
h_1^{(d-1)} &0 &\cdots &0\\
0 & h_2^{(d-1)} &\cdots &0\\
\vdots &\vdots &\ddots &\vdots\\
0 &0 &\cdots &h_{n_{d-1}}^{(d-1)}
\end{matrix}\right]
W^{(d-1)} \\
& \ \ \ \ \times \cdots \times
\left[\begin{matrix}
h_1^{(1)} &0 &\cdots &0\\
0 & h_2^{(1)} &\cdots &0\\
\vdots &\vdots &\ddots &\vdots\\
0 &0 &\cdots &h_{n_1}^{(1)}
\end{matrix}\right]
W^{(1)} \bfz\\
& =
\left[\begin{matrix}
h_1^{(d)} \left(w_1^{(d)}\right)^T \\
h_2^{(d)} \left(w_2^{(d)}\right)^T \\
\vdots\\
h_{n}^{(d)} \left(w_n^{(d)}\right)^T
\end{matrix}\right]
\left[\begin{matrix}
h_1^{(d-1)} \left(w_1^{(d-1)}\right)^T \\
h_2^{(d-1)} \left(w_2^{(d-1)}\right)^T \\
\vdots\\
h_{n_{d-1}}^{(d-1)} \left(w_{n_{d-1}}^{(d-1)}\right)^T
\end{matrix}\right]
\cdots
\left[\begin{matrix}
h_1^{(1)} \left(w_1^{(1)}\right)^T \\
h_2^{(1)} \left(w_2^{(1)}\right)^T \\
\vdots\\
h_{n_1}^{(1)} \left(w_{n_1}^{(1)}\right)^T
\end{matrix}\right]
\bfz,
\end{align*}
where ${h^{(i)}_j\in[h,1].}$

Denote the scaled matrix by
$$
P^{(1)} =
\left[\begin{matrix}
h_1^{(1)} \left(w_1^{(1)}\right)^T \\
h_2^{(1)} \left(w_2^{(1)}\right)^T \\
\vdots\\
h_{n_1}^{(1)} \left(w_{n_1}^{(d)}\right)^T
\end{matrix}\right]
,\cdots
,
P^{(d)} = \left[\begin{matrix}
h_1^{(d)} \left(w_1^{(d)}\right)^T \\
h_2^{(d)} \left(w_2^{(d)}\right)^T \\
\vdots\\
h_{n}^{(d)} \left(w_n^{(d)}\right)^T
\end{matrix}\right],
$$
then we know that $P^{(i)}, i=1,\cdots,d$ have independent Gaussian entries and the generative model becomes
\begin{align}\label{Eq:GMLeakyReLUNNScaling}
G(\bfz) = P^{(d)}P^{(d-1)}\cdots P^{(1)}\bfz.
\end{align}

From Lemma \ref{Lem:FullRankAndZeroEntries}, we only need to show that every $r=n-(2l+1)$ rows of $P^{(d)}P^{(d-1)}\cdots P^{(1)}$ have full rank $k$. {{Following}} the arguments in proving Lemma \ref{Lem:LinearNNArbitraryLayers}, we can show that every $r=n-(2l+1)$ rows of $P^{(d)}P^{(d-1)}\cdots P^{(1)}$ has full rank $k$. Thus we can arrive at our conclusion.

$\hfill\square$

Now we can get the recovery guarantee for generator implemented by arbitrary layer neural network with leaky ReLU activation functions as stated in Theorem \ref{Thm:NonlinearNNRecoveryGuarantee}.

\begin{theorem}\label{Thm:NonlinearNNRecoveryGuarantee}

Let the generator $G(\cdot)$ be implemented by a neural network described in Lemma \ref{Lem:LeakyReLUNNArbitraryLayers}. Let the true signal $\bfx_0$ and its corresponding low dimensional signal $\bfz_0$, the outlier signal $\bfe$ and observation $\bfy$ be the same as in Theorem \ref{Thm:L0MinimizationRecoveryCondition}. Let the measurement matrix $\mathcal{M}\in\mathbb{R}^{m\times n}$ be a random matrix with all entries {{iid random}} according to the Gaussian distribution and {{$\min(m,n)> l$}}. Then with high probability, the $\bfz_0$ can be recovered by solving
$$
\min_{\bfz\in\mathbb{R}^n} \|\mathcal{M}G(\bfz) - \y\|_0.
$$

\end{theorem}

{\emph Proof:} Following the argument in Lemma \ref{Lem:LeakyReLUNNArbitraryLayers}, we have
$$
\mathcal{M}G(\bfz) = \mathcal{M}P^{(d)} P^{(d-1)} \cdots P^{(1)}\bfz.
$$

Similar to the case where the generator is implemented by a linear neural network, the matrix $\mathcal{M}$ can be treated as the weight matrix in the $(d+1)$-th layer, and $P^{(i)}$ is the weight matrix in the $i$-th layer, and the activation functions are identity maps. {{Then according to Lemma \ref{Lem:LinearNNArbitraryLayers}, every $n-(2l+1)$ rows of {$P=P^{(d)}P^{(d-1)}\cdots P^{(1)}$} will have full rank $k$, and Lemma \ref{Lem:FullRankAndZeroEntries} implies that the $P(z-z_0)$ can have at most $n-(2l+1)$ zero entries or at least $(2l+1)$ nonzero elements. Finally Theorem \ref{Thm:L0MinimizationRecoveryCondition} gives that: with high probability, the $\bfz_0$ can be recovered by solving
$$
\min_{\bfz\in\mathbb{R}^n} \|\mathcal{M}G(\bfz) - \y\|_0.
$$
}}
$\hfill\square$

\section{Numerical Results}\label{Sec:NumericalResults}
In this section, we present experimental results to validate our theoretical analysis. We first { introduce} the datasets and generative model used in our experiments. Then, we { present simulation} results { verifying} our proposed { outlier detection algorithms.}
	
{We use two datasets MNIST\cite{Lecun:1998} and CelebFaces Attribute (CelebA) \cite{Liu:2015} in our experiments. For generative models, we use variational auto-encoder (VAE) \cite{Goodfellow:2014,Kingma:2013} and deep convolutional generative adversarial networks (DCGAN) \cite{Goodfellow:2014,Goodfellow:2016}.}

\subsection{MNIST and VAE}

The MNIST database contains approximately 70,000 images of handwritten digits which are divided into two separate sets: training and test sets. The training set has 60,000 examples and {the} test set has 10,000 examples. Each image {is of size} 28 $\times$ 28 where each pixel { is} either value of '1' or '0'.
	
For the VAE, we { adopt the pre-trained VAE generative model in \cite{bora_compressed_2017} for MNIST dataset.
The VAE system consists of an encoding and a decoding networks in the inverse architectures.} Specifically, the encoding network has fully connected structure with input, output and two hidden layers. The input has the vector size of 784 elements, the output produces the latent vector having the size of 20 elements. Each hidden layer possesses 500 neurons. The fully connected network creates 784 -- 500 -- 500 -- 20 system. Inversely, the decoding network consists three fully connected layer converting from latent representation space to image space, i.e., 20--500--500--784 network. { In the training process, the decoding network is trained in such a way to produce the similar distribution of original signals. Training dataset of 60,000 MNIST image samples was used to train the VAE networks. The training process was conducted by Adam optimizer \cite{Kingma:2013} with batch size 100 and learning rate of 0.001.}

When the VAE is well-trained, we exploit the decoding neural network as a generator to produce input-like signal $x' = G(\z)\in\mathbb{R}^n$. This processing maps a low dimensional signal $\z\in\mathbb{R}^k$ to a high dimensional signal $G(\z)\in\mathbb{R}^n$ such that $\|G(\z) - \x\|_2$ can be small, where the $k$ is set to be $20$ and $n=784$.

Now, we solve the compressed sensing problem to find the optimal latent vector $\hat{\bf z}$ and its corresponding representation vector $G(\z)$ by $\ell_1$-norm and { $\ell_2$-norm minimizations} which are explained as follows. For $\ell_1$ minimization, we consider both the case where we solve
\begin{align}\label{Defn:UnSquaredL1MinimizationViaADMM}
\min_{\z\in\mathbb{R}^k} \|\mathcal{M}G(\z) - \y\|_1
\end{align}
by the ADMM algorithm introduced in Section \ref{Sec:ADMMAlgorithm}, and the case where we solve
\begin{align}\label{Defn:SquaredL1MinimizationViaGD}
\min_{\z\in\mathbb{R}^k} \|\mathcal{M}G(\z) - \y\|_1^2
\end{align}
by gradient descent (GD) algorithm \cite{Barzilai88}. {  We note that the optimization problem in (\ref{Defn:UnSquaredL1MinimizationViaADMM}) is non-differentiable w.r.t $\bf z$, thus, it cannot be solved by GD algorithm. The optimization problem in (\ref{Defn:SquaredL1MinimizationViaGD}), however, can be solved in practice using GD algorithm. It is because (\ref{Defn:SquaredL1MinimizationViaGD}) is only non-differentiable at a finite number of points. Therefore, (\ref{Defn:SquaredL1MinimizationViaGD}) can be effectively solved by a numerical GD solver. }{ In what follows,} we will simply refer to the former case (\ref{Defn:UnSquaredL1MinimizationViaADMM}) as { $\ell_1$ ADMM} and the later case (\ref{Defn:SquaredL1MinimizationViaGD}) as { $(\ell_1)^2$ GD}. For $\ell_2$ minimization, we { solve the following optimization problems:},
\begin{align}\label{Defn:UnRegularizedL2Minimization}
\min_{\z\in\mathbb{R}^k} \|\mathcal{M}G(\z) - \y\|_2^2,
\end{align}
and
\begin{align}\label{Defn:RegularizedL2Minimization}
\min_{\z\in\mathbb{R}^k} \|\mathcal{M}G(\z) - \y\|_2^2 + \lambda \|\z\|_2^2.
\end{align}
Both (\ref{Defn:UnRegularizedL2Minimization}) and (\ref{Defn:RegularizedL2Minimization}) are solved by \cite{Barzilai88}. {We} will refer to the former case (\ref{Defn:UnRegularizedL2Minimization}) as { $(\ell_2)^2$ GD}, and the later (\ref{Defn:RegularizedL2Minimization}) as { $(\ell_2)^2$ GD + reg}. The regularization parameter $\lambda$ in (\ref{Defn:RegularizedL2Minimization}) is set to be $0.1$.

\subsection{CelebA and DCGAN}

The CelebFaces Attribute dataset consists of over 200,000 celebrity images. We first resize the image samples to $64\times 64$ RGB images, and each image will have {in total} $12288$ pixels with values from $[-1,1]$. { We adopt the pre-trained deep convolutional generative adversarial networks (DCGAN) in \cite{bora_compressed_2017} to conduct the experiments on CelebA dataset}. Specifically, DCGAN consists of a generator and a discriminator. Both generator and discriminator have the structure of one input layer, one output layer and 4 convolutional hidden layers. We map the vectors from latent space of size $k$ = 100 elements to signal space vectors having the size of $n=64\times64\times3= 12288$ elements. 

In the training process, { both the generator and the discriminator were trained}. While training the discriminator to identify the {fake} images generated from the generator and ground truth images, the generator is also trained to increase the quality of its fake images so that the discriminator cannot identify the generated samples and ground truth image. This idea {is} based on Nash equilibrium \cite{Nash48} of game theory.
	
{ The DCGAN was trained with 200,000 images from CelebA dataset.  We further use additional 64 images for testing our CS.} The training process was conducted by the Adam optimizer \cite{Kingma:2014} with learning rate 0.0002, 
momentum $\beta_1 = 0.5$ and batch size of 64 images.

Similarly, we solve our CS problems in (\ref{Defn:UnSquaredL1MinimizationViaADMM}), (\ref{Defn:SquaredL1MinimizationViaGD}), (\ref{Defn:UnRegularizedL2Minimization}) and (\ref{Defn:RegularizedL2Minimization}) by { $\ell_1$ ADMM, $(\ell_1)^2$ GD}, { $(\ell_2)^2$ GD} and { $(\ell_2)^2$ GD + reg} algorithms, respectively. The regularization parameter $\lambda$ in (\ref{Defn:RegularizedL2Minimization}) is set to be $0.001$. Besides, we also solve the outlier detection problem by Lasso on the images in both the discrete cosine transformation domain (DCT) \cite{Ahmed:74} and the wavelet transform domain (Wavelet) \cite{Daubechies:88}.

\subsection{Experiments and Results}

\subsubsection{Reconstruction with various numbers of measurements}

The theoretical result in \cite{bora_compressed_2017} showed that a random Gaussian measurement $\cal M$ is effective for signal reconstruction in CS. Therefore, for evaluation, we set $\cal M$ as a random matrix with i.i.d. Gaussian entries with zero mean and { standard deviation of 1}. Also, every element of noise vector $\eta$ is an i.i.d. Gaussian random variable.
In our experiments, we carry out the performance comparisons between our proposed $\ell_1$ minimization (\ref{Defn:UnSquaredL1MinimizationViaADMM}) with ADMM algorithm (referred as { $\ell_1$ ADMM}) approaches, { $(\ell_1)^2$} minimization (\ref{Defn:SquaredL1MinimizationViaGD}) with GD algorithm (referred as { $(\ell_1)^2$ GD}), regularized { $(\ell_2)^2$ minimization} (\ref{Defn:RegularizedL2Minimization}) with GD algorithm (referred as{ $(\ell_2)^2$ GD + reg}), $(\ell_2)^2$ minimization (\ref{Defn:UnRegularizedL2Minimization}) with GD algorithm (referred as { $(\ell_2)^2$ GD}), Lasso in DCT domain, and Lasso in wavelet domain \cite{bora_compressed_2017}. We use the reconstruction error as our performance metric, which is defined as: $\rm{error} = \| \hat{\bf x} - {\bf x}^* \|_2^2$, where $\hat{\bf x}$ is an estimate of ${\bf x}^*$ returned by the algorithm. 

For MNIST, we set the standard deviation of the noise vector so that $\sqrt{\mathbb{E}[\|\bf \eta \|^2]} = 0.1$. We conduct $10$ random restarts with $1000$ steps per restart and pick the reconstruction with best measurement error. For CelebA, we set the standard deviation of entries in the noise vector so that $\sqrt{\mathbb{E}[\|\bf \eta \|^2]} = 0.01$. We conduct $2$ random restarts with $500$ update steps per restart and pick the reconstruction with best measurement error.

Fig.\ref{fig:amnist-reconstr-l1} and Fig.\ref{fig:amnist-reconstr-l2} show the performance of CS system in the presence of 3 outliers. We compare the reconstruction error versus {the} number of measurements for the $\ell_1$-minimization based methods with ADMM (\ref{Defn:UnSquaredL1MinimizationViaADMM}) and GD (\ref{Defn:SquaredL1MinimizationViaGD}) algorithms in Fig.\ref{fig:amnist-reconstr-l1}. In Fig.\ref{fig:amnist-reconstr-l2}, we plot the reconstruction error versus number of measurements for $\ell_2$-minimization based methods with and without regularization in (\ref{Defn:UnRegularizedL2Minimization}) and (\ref{Defn:RegularizedL2Minimization}), respectively. {The outliers' values and positions are randomly {{generated}}. To generate the outlier vector, we first create a vector ${\bf{e}} \in {\R^m}$ having all zero elements. Then, we randomly generate $l$ integers in the range $[1,m]$ indicating the positions of outlier in vector $\bf e$. Now, for each generated outlier position, we assign a large random value in the range [5000, 10000]. The outlier vector $\bf e$ is then added into CS model as in (\ref{Defn:MeasurementWithOutlier}).} As we can see in Fig.\ref{fig:amnist-reconstr-l1} when $\ell_1$ minimization algorithms { ($\ell_1$ ADMM and $(\ell_1)^2$ GD)} were used, the reconstruction errors fast converge to low values as the number of measurements increases. On the other hand, the $\ell_2$-minimization based recovery do not {{work well}} as can be seen in  Fig.\ref{fig:amnist-reconstr-l2} even when number of measurements increases to a large quantity. One observation can be made from Fig.\ref{fig:amnist-reconstr-l1} is that after $100$ measurements, our algorithm's performance saturates, higher number of measurements does not enhance the reconstruction error performance. This is because of the limitation of VAE architecture.

Similarly, we conduct our experiments with CelebA dataset  using {the} DCGAN generative model. In Fig. \ref{fig:acelebA-reconstr-l1}, we show the reconstruction error {{change}} as we increase the number of measurements both for  $\ell_1$-minimization based with {the $\ell_1$ ADMM and $(\ell_1)^2$ GD algorithms}. In Fig.~\ref{fig:acelebA-reconstr-l2}, we compare reconstruction errors of $\ell_2$-minimization using GD algorithms and Lasso { with the DCT and wavelet bases}. We observed that {in} the presence of outliers, $\ell_1$-minimization based methods outperform the $\ell_2$-minimization based  algorithms and methods using Lasso.

We plot the sample reconstructions by Lasso and our algorithms in Fig. \ref{fig:mnist_sample_m100}. We observed that { $\ell_1$ ADMM and $(\ell_1)^2$ GD} approaches are able to reconstruct the images with only as few as 100 measurements while conventional Lasso and $\ell_2$ minimization algorithm are unable to do the same when number of outliers is as small as 3.

For CelebA dataset, we first show the {{reconstruction performance}} when the number of outlier is 3 and {{the}} number of measurements is 500. We plot the reconstruction results using {$\ell_1$ ADMM and $(\ell_1)^2$ GD} algorithms with DCGAN in Fig. \ref{fig:celebA-reconstr_l1_o3}. Lasso DCT and Wavelet, { $(\ell_2)^2$ GD and $(\ell_2)^2$ GD + reg} with DCGAN reconstruction results are showed in Fig. \ref{fig:celebA-reconstr_l2_o3}. Similar to the result in MNIST set, the proposed  $\ell_1$-minimization based algorithms with DCGAN perform better in the presence of outliers. This is because $\ell_1$-minimization-based approaches can successfully eliminate the outliers while $\ell_2$-minimization-based methods do not.

We show {{further}} results for both two datasets with various numbers of measurements in Fig. \ref{fig:more-mnist-reconstr_l1}, \ref{fig:more-mnist-reconstr_l2}, \ref{fig:more-celebA-reconstr1_l1}, and \ref{fig:more-celebA-reconstr1_l2}. Specifically, Fig. \ref{fig:more-mnist-reconstr_l1} shows MNIST reconstruction results using $\ell_1$-minimization based algorithms, and Fig. \ref{fig:more-mnist-reconstr_l2} plots MNIST reconstruction results using $\ell_2$-minimization based algorithm and Lasso, respectively. Fig. \ref{fig:more-celebA-reconstr1_l1}, and \ref{fig:more-celebA-reconstr1_l2} display {{sample}} results for CelebA recovery with different algorithms.

\subsubsection{Reconstruction with different numbers of outliers}

 Now, we evaluate the recovery performance of CS system under different numbers of outliers. We first fix the noise levels for MNIST and CelebA as the same as in the previous experiments. Then, for each dataset, we vary the number of outliers from 5 to 50, and measure the reconstruction error per pixel with various numbers of measurements. In these evaluations, we use $\ell_1$-minimization based algorithms {($\ell_1$ ADMM and $(\ell_1)^2$ GD)} for outlier detection and image reconstruction.

For MNIST dataset, we plot the reconstruction error performance for 5, 10, 25 and 50 outliers, respectively, in Fig. \ref{fig:mnist-reconstr-o5to50}. One can be seen from Fig. \ref{fig:mnist-reconstr-o5to50} that as the number of outliers increases, larger number of measurements are needed to guarantee {{successful}} image recovery. Specifically, we need at least 25 measurements to lower error rate to below 0.08 per pixel when there are 5 outliers in data, while the number of measurements should be tripled to obtain the same performance { in} the presence of 50 outliers. We show the image reconstruction results using ADMM and GD algorithms with different { numbers} of outliers in Fig. \ref{fig:mnist-sample-o1to25}.

 Next, for { the CelebA} dataset, we plot the reconstruction error performance for 5, 10, 25 and 50 outliers in Fig. \ref{fig:celebA-reconstr-o5to50}. We compare the image reconstruction results using { $\ell_1$ ADMM and $(\ell_1)^2$ GD} algorithms in Fig. \ref{fig:celebA-sample-o1to25-a}.

\section{{ Conclusions} and Future Directions}\label{Sec:ConclusionsandFutureDirections}

{{

In this paper, we investigated {{solving}} the outlier detection problems via a generative model approach. This new approach outperforms the $\ell_2$ minimization and traditional Lasso in both the DCT domain and Wavelet domain. The iterative alternating {{direction}} method of multipliers we proposed can {{efficiently solve the proposed nonlinear $\ell_1$ norm-based outlier detection formulation for generative model.}} Our theory shows that for both the linear neural networks and nonlinear neural networks with arbitrary {{number of}} layers, as long as they satisfy certain mild conditions, then with high probability, {{one can correctly detect the outlier signals based on generative models.}}

}}

\bibliographystyle{unsrt}
\bibliography{CS_ref}

\begin{figure*}
	\newsavebox\myboxone
	\savebox{\myboxone}{\includegraphics[width=0.48\textwidth]{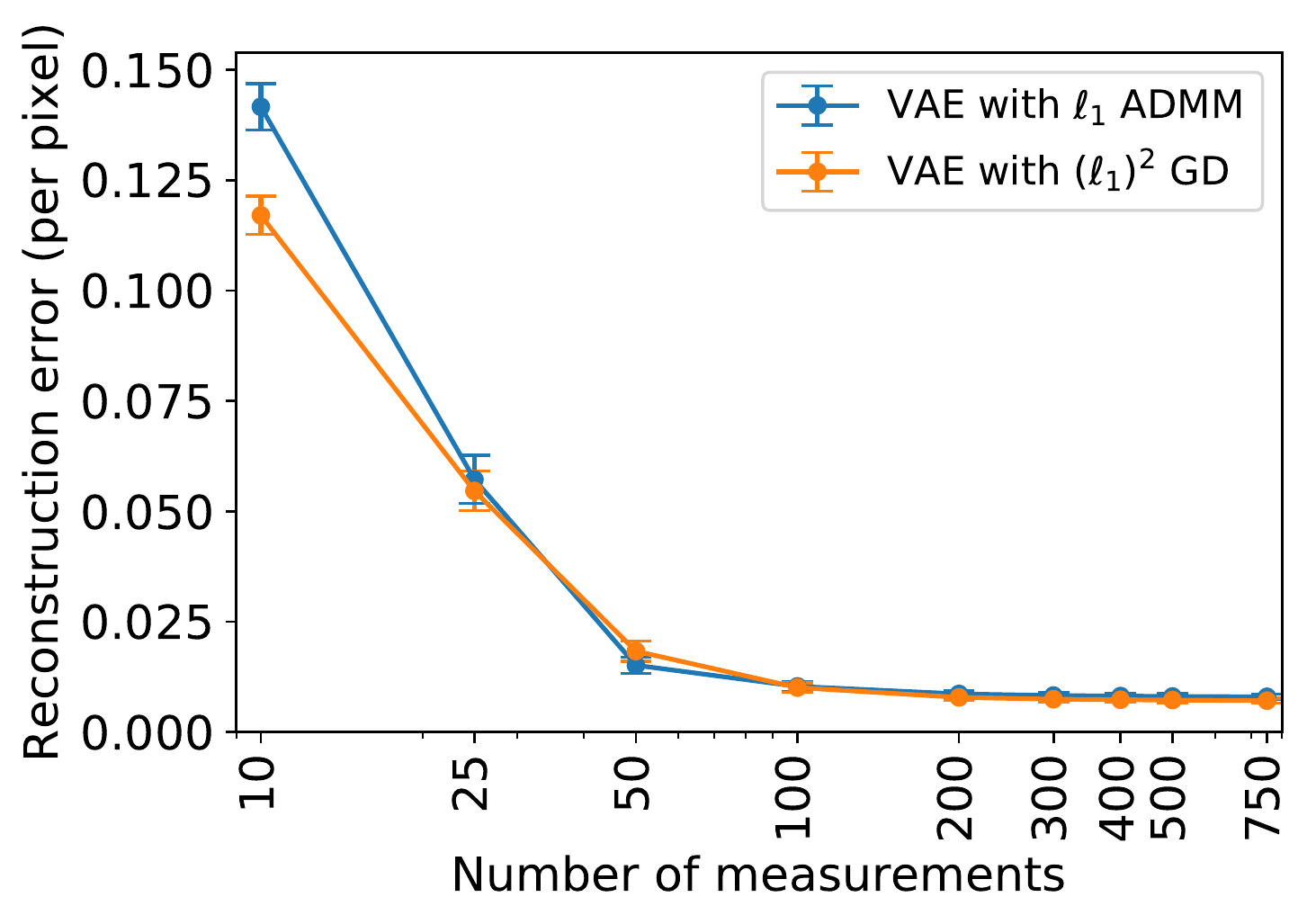}}
	\begin{subfigure}[t]{0.5\textwidth}
		\usebox{\myboxone}
		\vspace*{-3mm}
		\caption{$\ell_1$- minimization algorithms}
		\label{fig:amnist-reconstr-l1}
	\end{subfigure}\hfill%
	\begin{subfigure}[t]{0.5\textwidth}
		\vbox to \ht\myboxone{%
			\vfill
			\includegraphics[width=\textwidth]{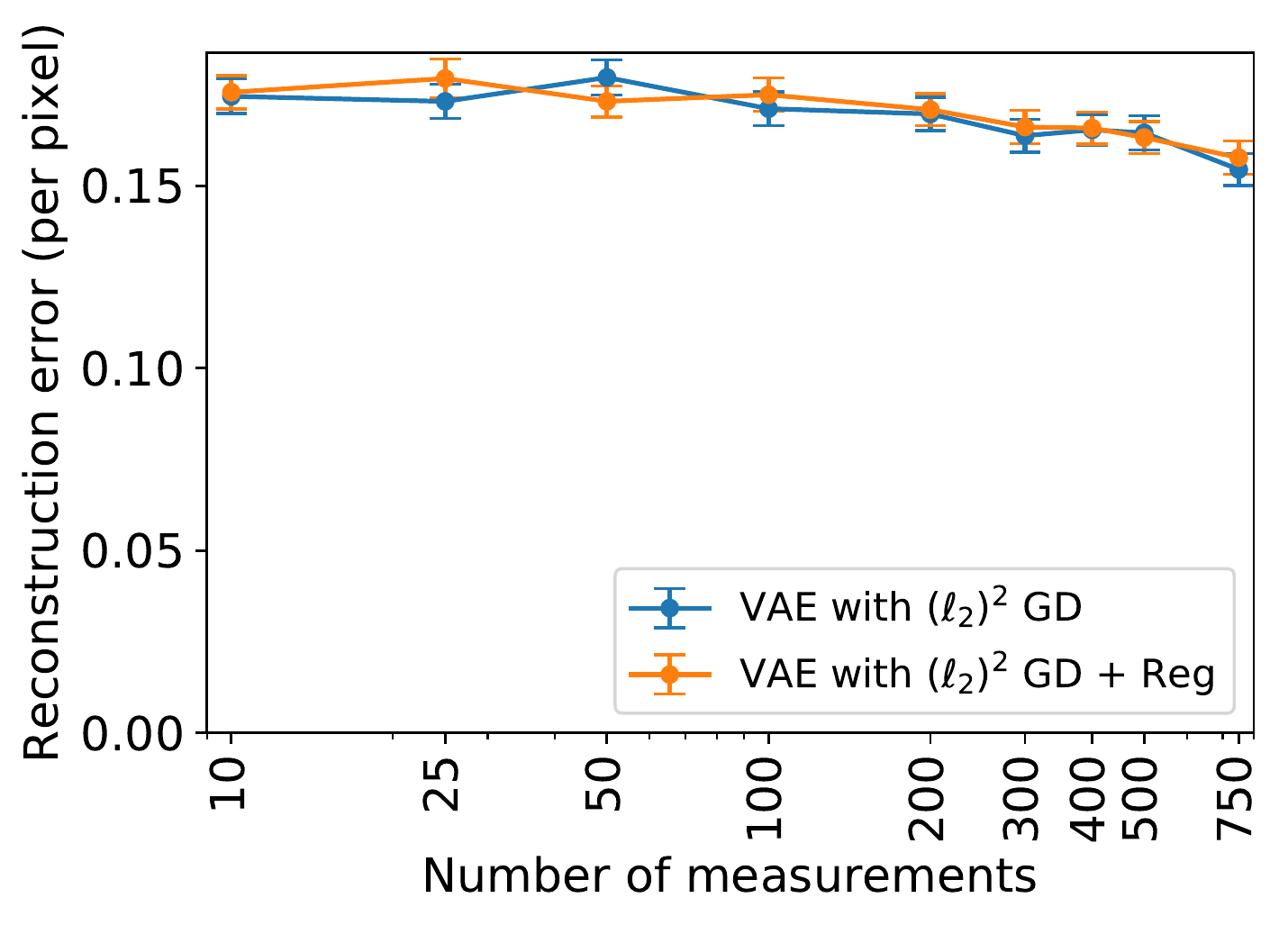}
		}
		\vspace*{2mm}
		\caption{$\ell_2$- minimization algorithms}
		\label{fig:amnist-reconstr-l2}
	\end{subfigure}
	\caption{MNIST results: We compare the reconstruction error performance { using} VAE among $\ell_1$ ADMM algorithm solving (\ref{Defn:UnSquaredL1MinimizationViaADMM}), {$(\ell_1)^2$ GD} algorithm solving (\ref{Defn:SquaredL1MinimizationViaGD}), {$(\ell_2)^2$ GD}  algorithm solving (\ref{Defn:UnRegularizedL2Minimization}) and {$(\ell_2)^2$ GD} + reg algorithm solving (\ref{Defn:RegularizedL2Minimization}). We plot the reconstructions error per pixel versus various numbers of measurements when the number of  outliers is 3.  The vertical bars indicate 95\% confidence intervals.}
	\vspace*{2mm}
	\newsavebox\myboxtwo
	\savebox{\myboxtwo}{\includegraphics[width=0.48\textwidth]{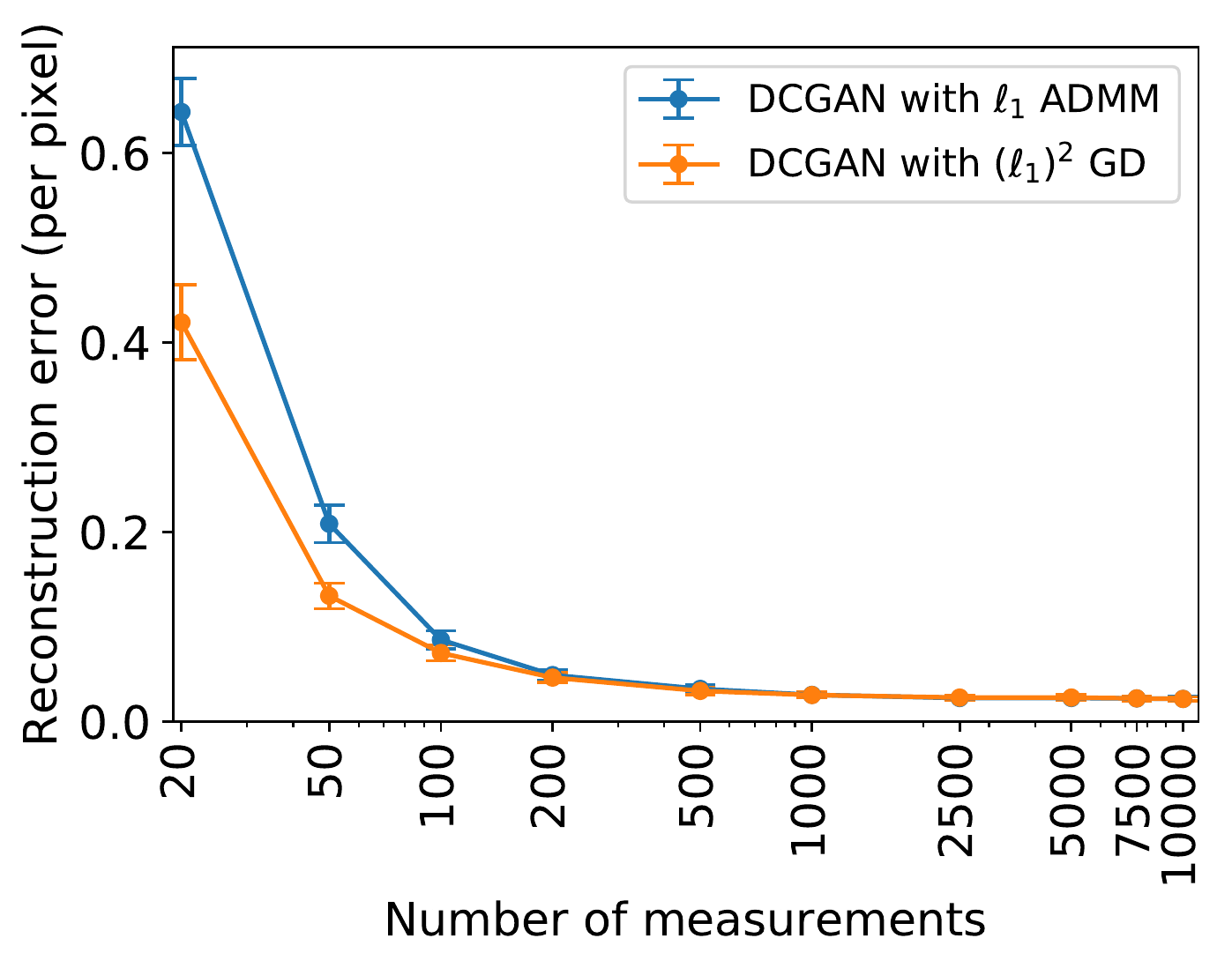}}
	\begin{subfigure}[t]{0.5\textwidth}
		\usebox{\myboxtwo}
		\vspace*{-3mm}
		\caption{$\ell_1$- minimization algorithms }
		\label{fig:acelebA-reconstr-l1}
	\end{subfigure}\hfill%
	\begin{subfigure}[t]{0.5\textwidth}
		\vbox to \ht\myboxtwo{%
			\vfill
			\includegraphics[width=\textwidth]{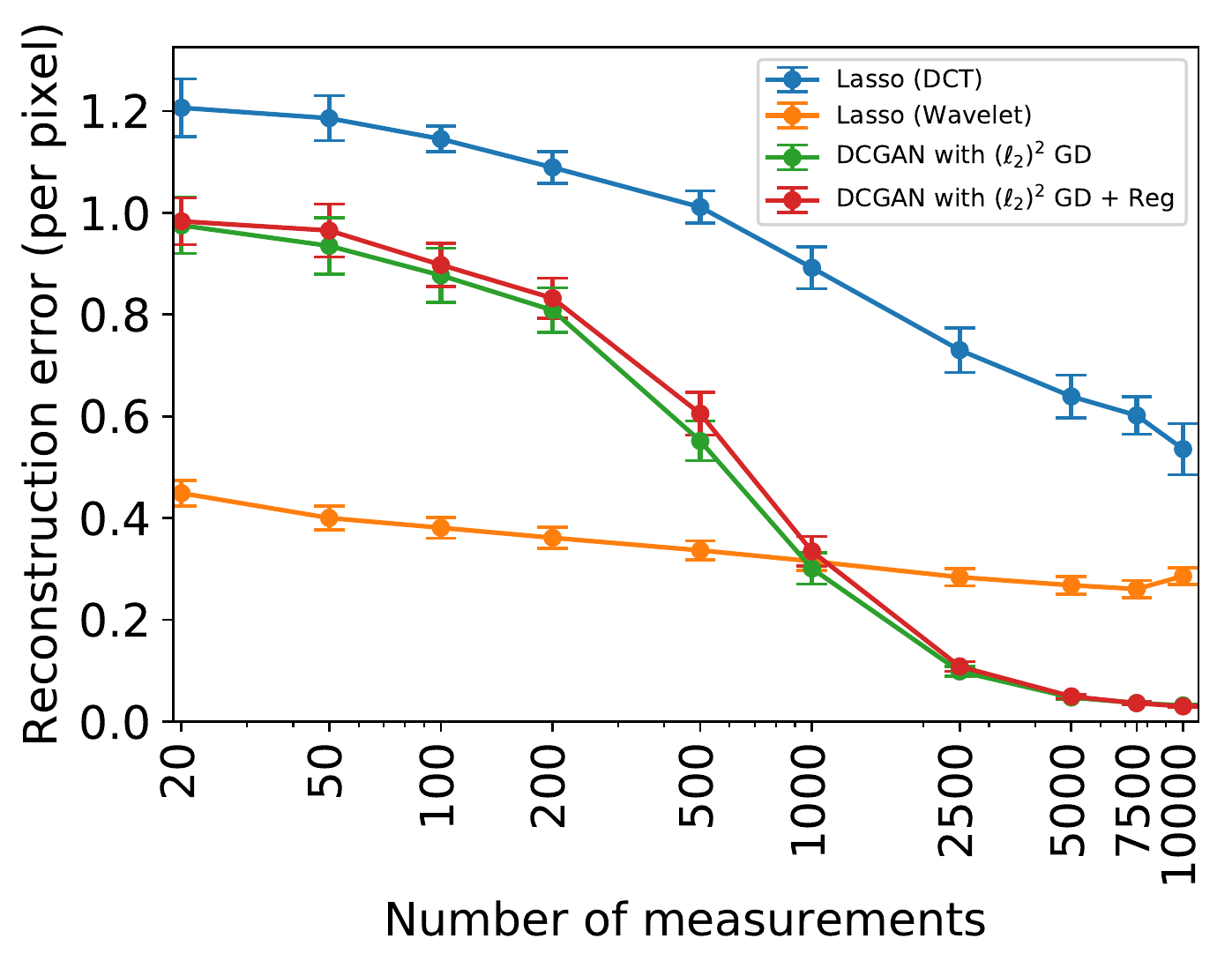}
		}
		\vspace*{2mm}
		\caption{$\ell_2$- minimization algorithm and lasso methods}
		\label{fig:acelebA-reconstr-l2}
	\end{subfigure}
	\caption{CelebA results: We compare the reconstruction error performance { using} DCGAN { among} $\ell_1$ ADMM algorithm solving (\ref{Defn:UnSquaredL1MinimizationViaADMM}), $(\ell_1)^2$ GD algorithm solving (\ref{Defn:SquaredL1MinimizationViaGD}), $(\ell_2)^2$ GD algorithm solving (\ref{Defn:UnRegularizedL2Minimization}), $(\ell_2)^2$ GD + regularization algorithm solving (\ref{Defn:RegularizedL2Minimization}), { and} Lasso with DCT and Wavelet bases methods. We plot the reconstructions error per pixel versus various numbers of measurements when the number of outliers is 3. The vertical bars indicate 95\% confidence intervals.}
\end{figure*}

\begin{figure*}
	\begin{subfigure}[t]{0.48\textwidth}
		\includegraphics[width=\textwidth]{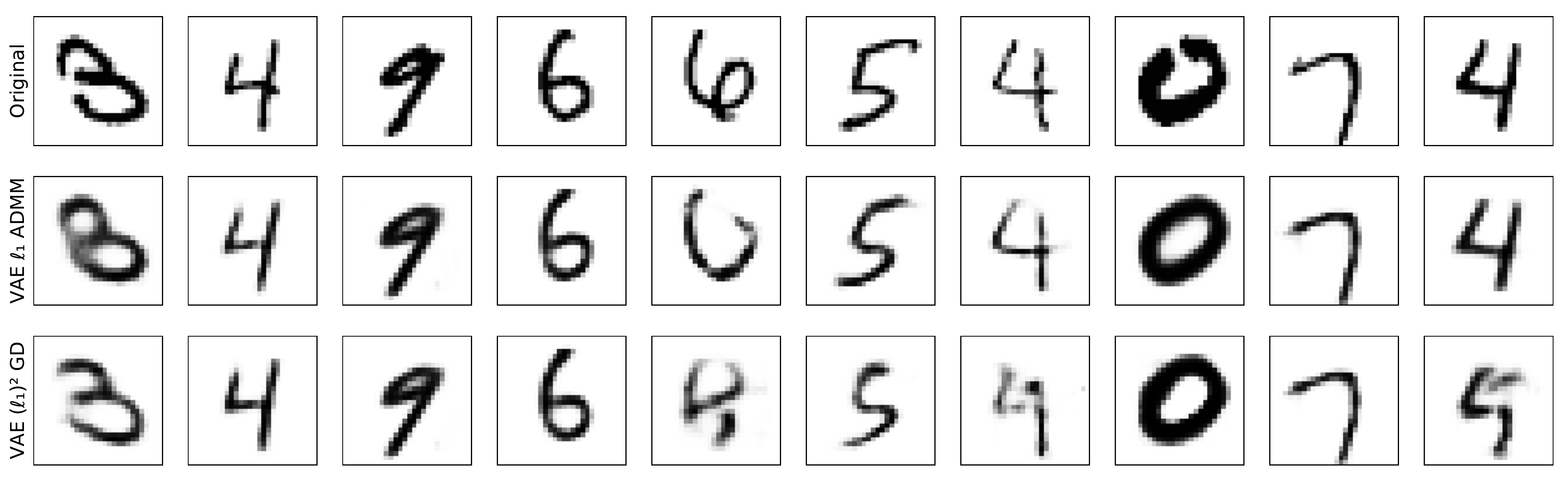}
		\caption{Top row: original images, middle row: reconstructions by VAE with $\ell_1$ ADMM algorithm, and bottom row: reconstructions by VAE with $(\ell_1)^2$ GD algorithm.}
		\label{fig:mnist_sample_l1}
	\end{subfigure}\hfill%
	\begin{subfigure}[t]{0.48\textwidth}
		\includegraphics[width=\textwidth]{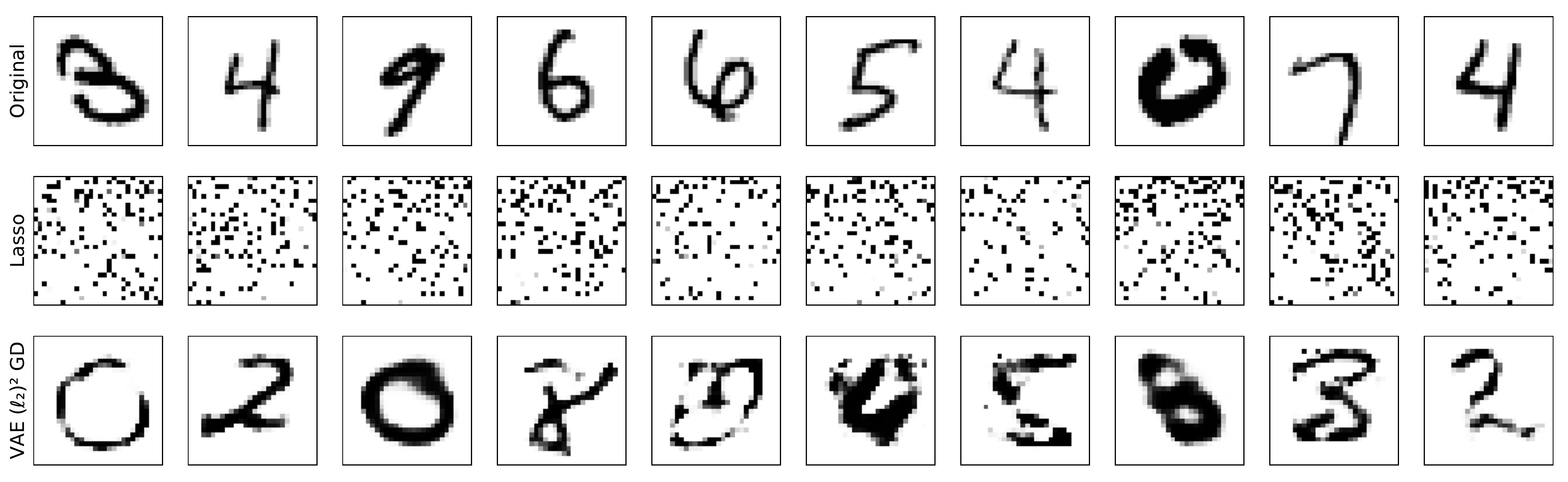}
		\caption{Top row: original images, middle row: reconstructions by Lasso, and bottom row: reconstructions by VAE with $(\ell_2)^2$ GD algorithm.}
		\label{fig:mnist_sample_l2}
	\end{subfigure}
	\caption{Reconstructions of MNIST samples when the number of measurements is 100 and the number of outliers is 3}
	\label{fig:mnist_sample_m100}
\end{figure*}

\begin{figure*}
	\centering
	\includegraphics[width=1\textwidth]{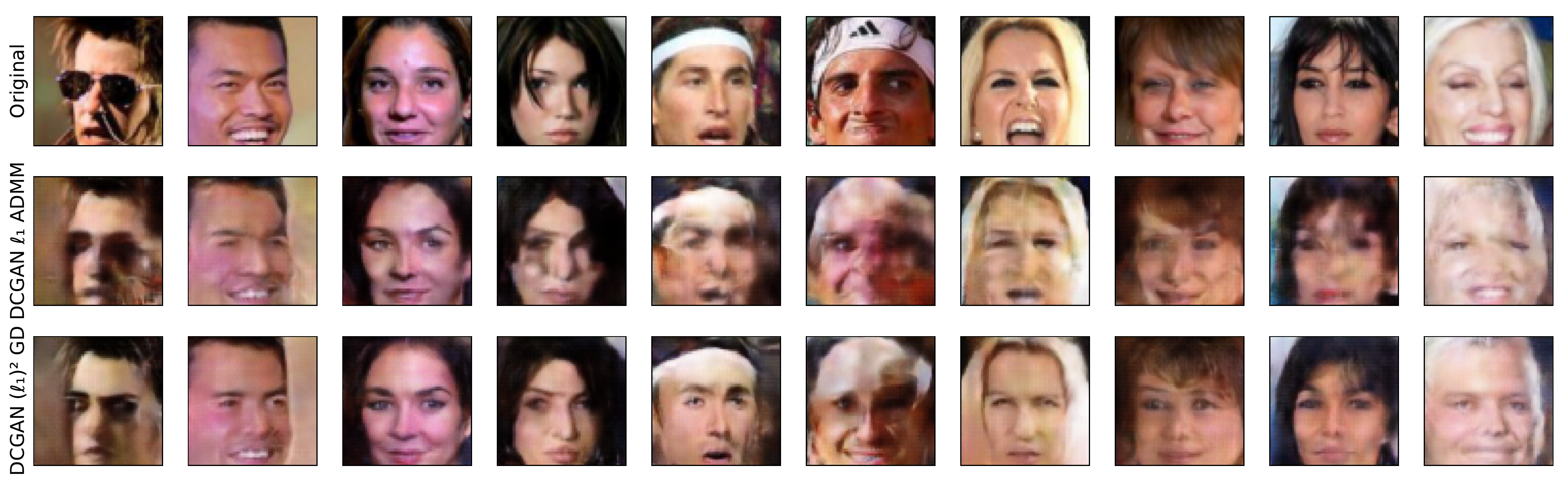}
	\caption{Reconstruction results of CelebA  samples when the number of measurements is 500 and the number of outliers is 3. Top row: original images, middle row: reconstructions by DCGAN with $\ell_1$ ADMM algorithm, and bottom row: reconstructions by DCGAN with $(\ell_1)^2$ GD algorithm.}
	\label{fig:celebA-reconstr_l1_o3}
\end{figure*}

\begin{figure*}
	\centering
	\includegraphics[width=1\textwidth]{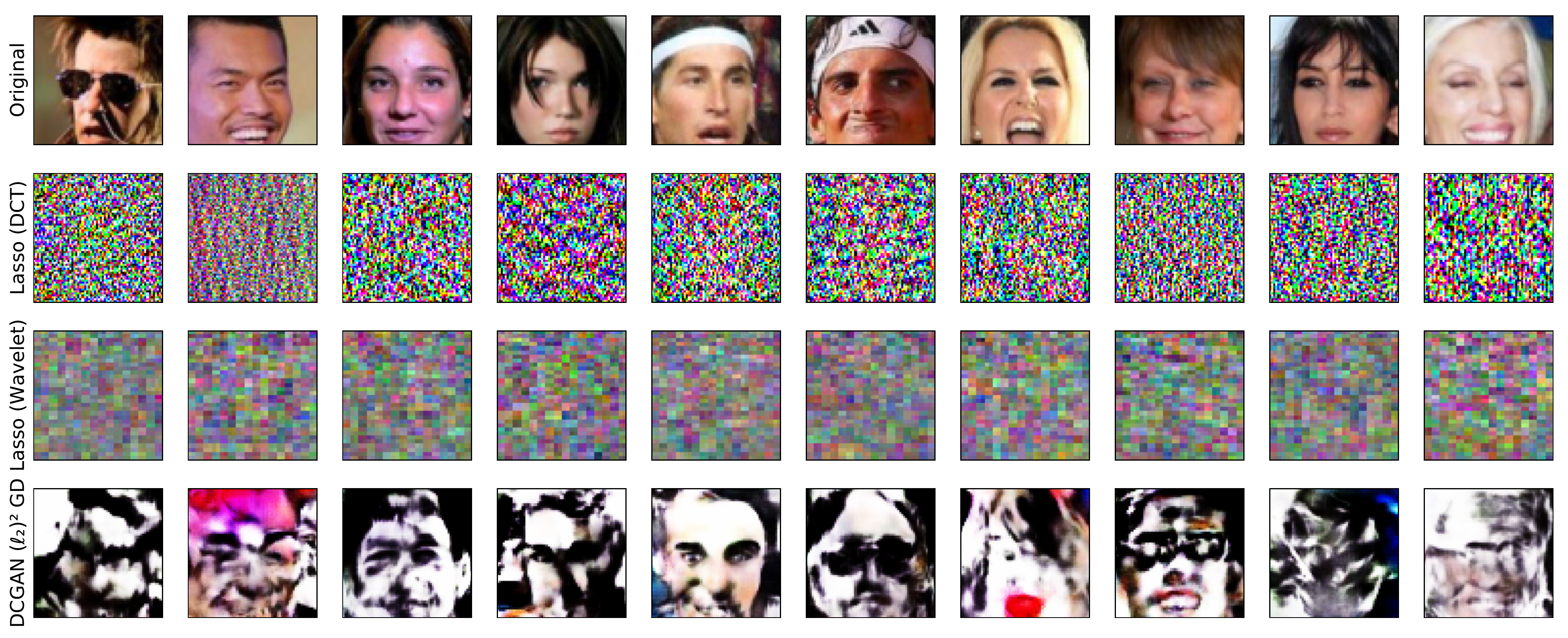}
	\caption{Reconstruction results of CelebA samples when the number of measurements is 500 and the number of outliers is 3. Top row: original images, middle row: reconstructions by Lasso with DCT basis, Third row: reconstructions by Lasso with wavelet basis, and last row: reconstructions by DCGAN with $(\ell_2)^2$ GD algorithm.}
	\label{fig:celebA-reconstr_l2_o3}
\end{figure*}

\begin{figure*}
	\centering
	\begin{subfigure}[]{1\textwidth}
		\includegraphics[width=\textwidth]{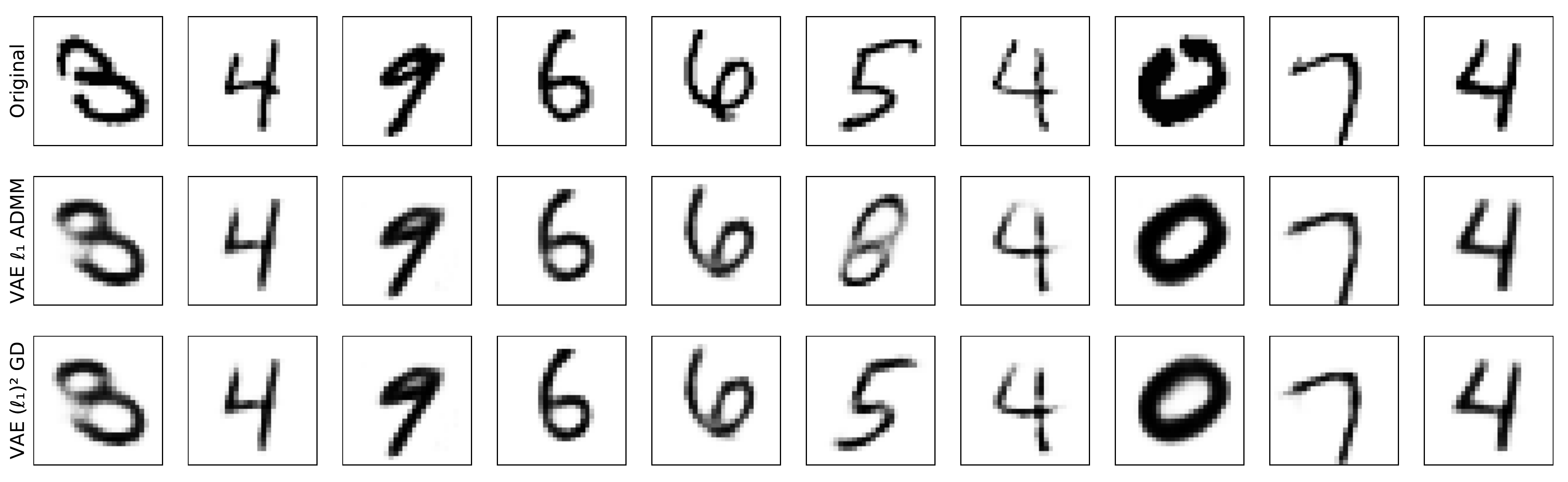}
		\caption{200 measurements}
	\end{subfigure}
	\begin{subfigure}[]{1\textwidth}
		\includegraphics[width=\textwidth]{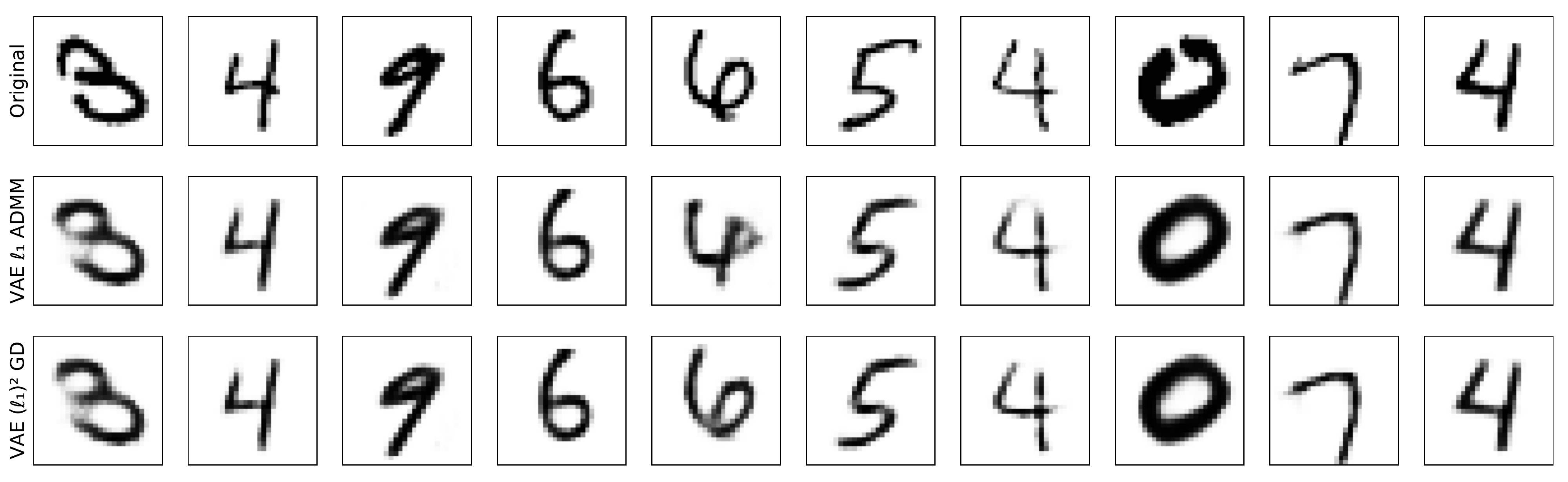}
		\caption{400 measurements}
	\end{subfigure}
	\begin{subfigure}[]{1\textwidth}
		\includegraphics[width=\textwidth]{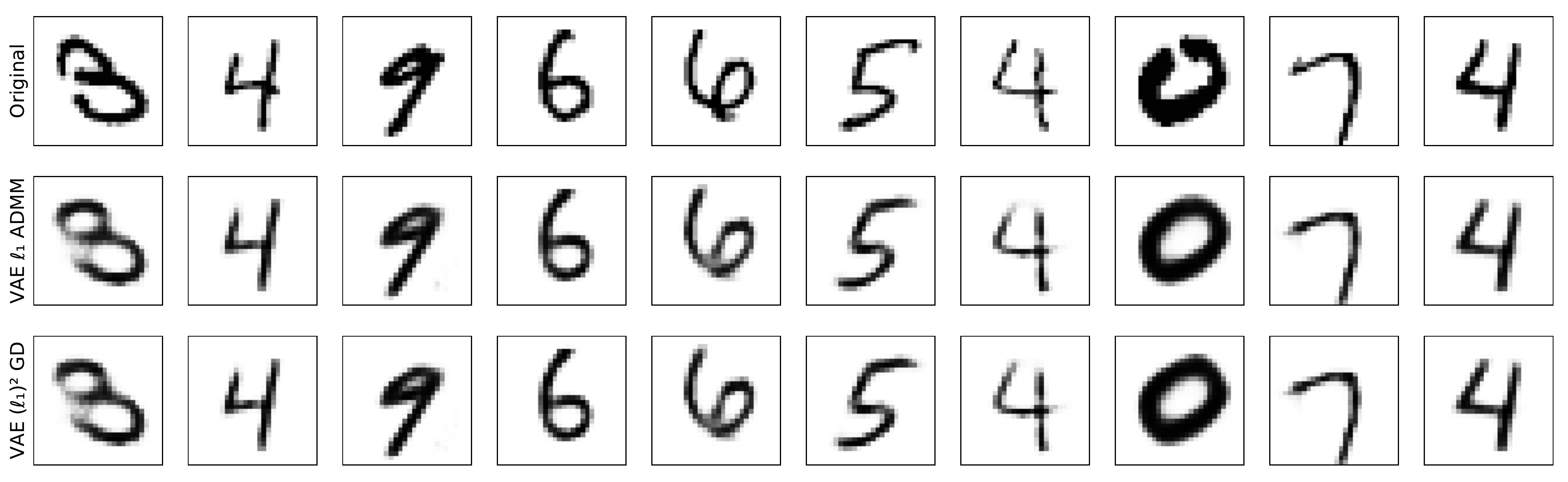}
		\caption{750 measurements}
	\end{subfigure}
	\caption{Reconstruction results of MNIST samples with different numbers of measurements when the number of outliers is 3.  Note that in each result, top row: original images, middle row: reconstructions by VAE with $\ell_1$ ADMM algorithm, and bottom row: reconstructions by VAE with $(\ell_1)^2$ GD algorithm.}
	\label{fig:more-mnist-reconstr_l1}
\end{figure*}

\begin{figure*}
	\centering
	\begin{subfigure}[]{1\textwidth}
		\includegraphics[width=\textwidth]{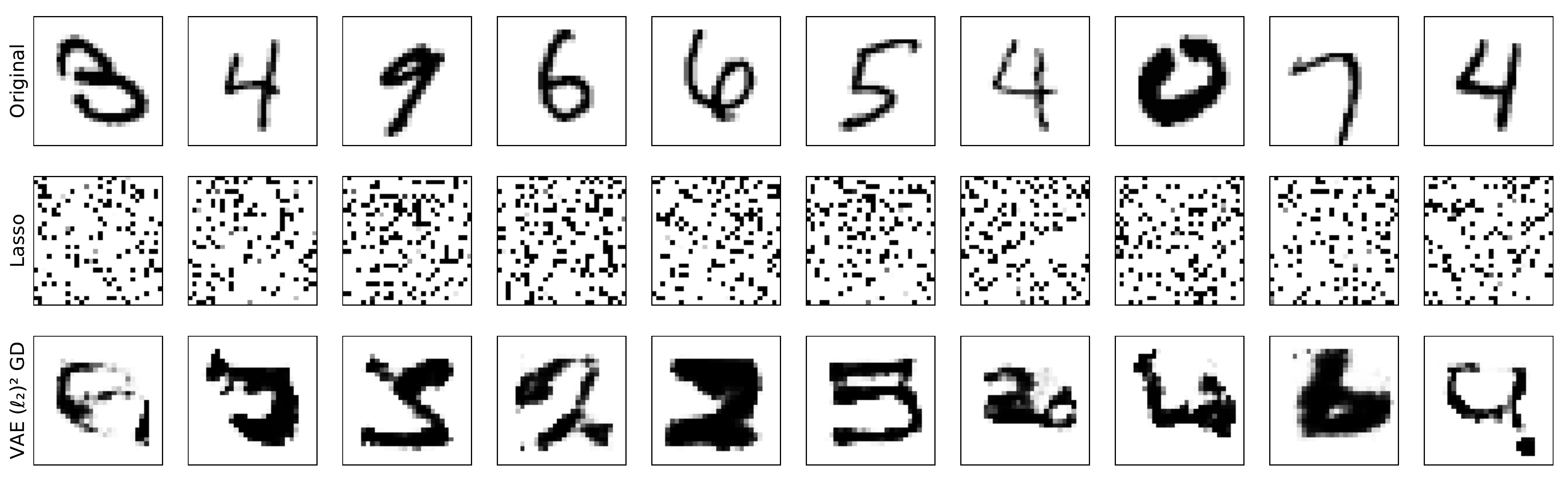}
		\caption{200 measurements}
	\end{subfigure}
	\begin{subfigure}[]{1\textwidth}
		\includegraphics[width=\textwidth]{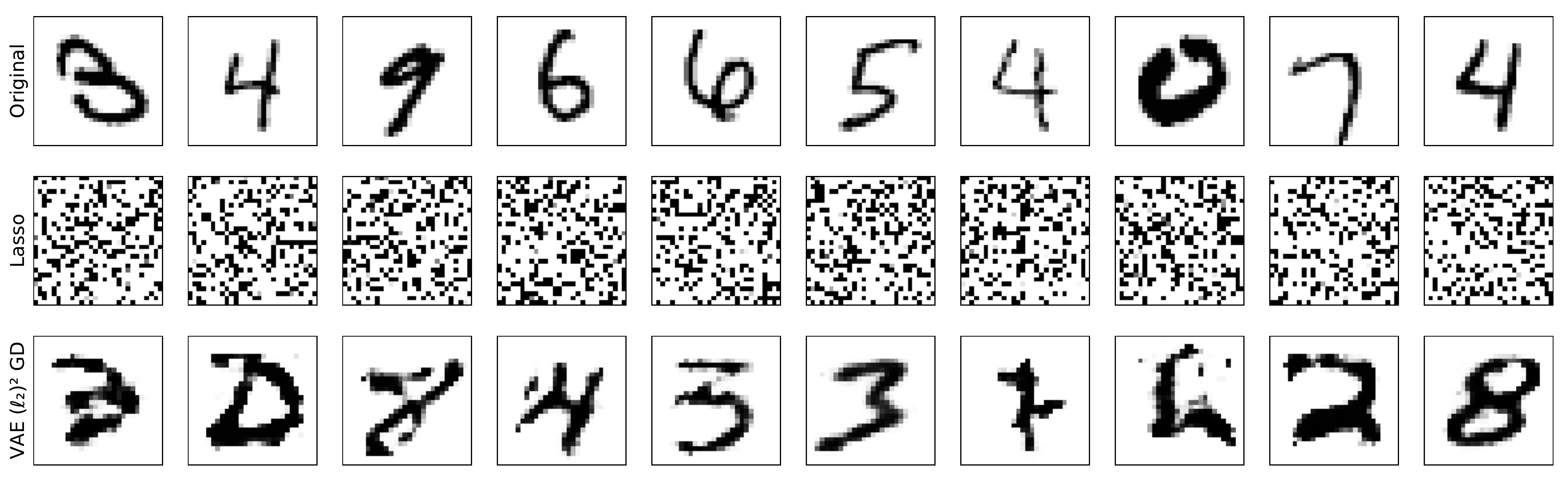}
		\caption{400 measurements}
	\end{subfigure}
	\begin{subfigure}[]{1\textwidth}
		\includegraphics[width=\textwidth]{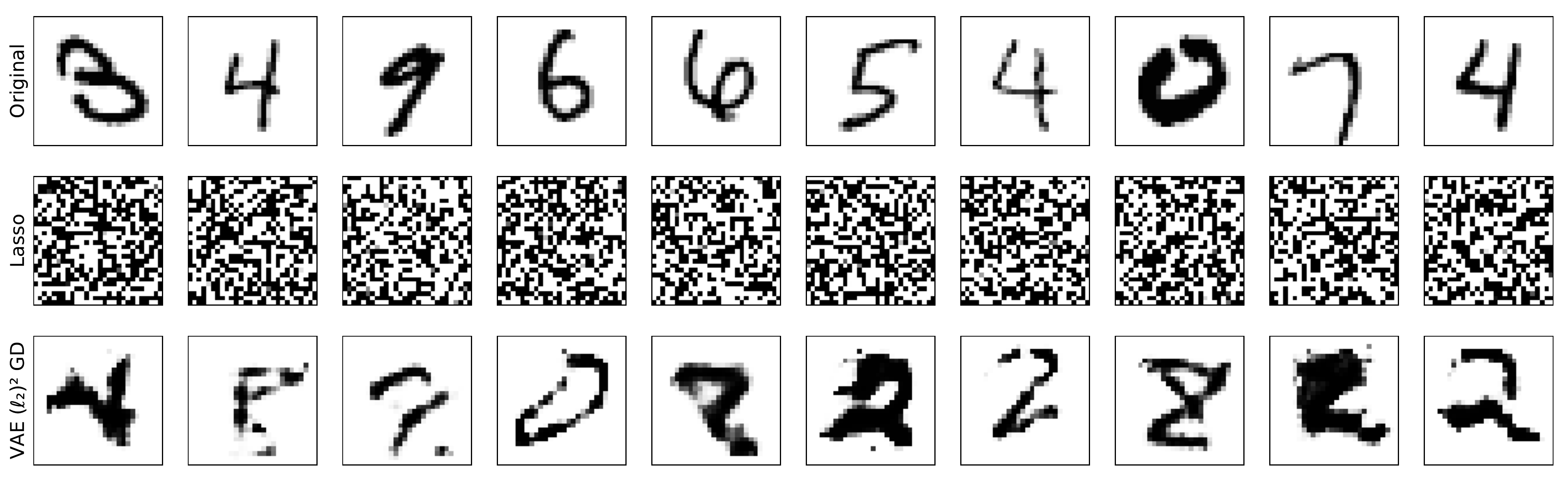}
		\caption{750 measurements}
	\end{subfigure}
	\caption{Reconstruction results of MNIST samples with different numbers of measurements when the number of outliers is 3 (continue). Note that in each result, top row: original images, middle row: reconstructions by Lasso, and bottom row: reconstructions by VAE with $(\ell_2)^2$ GD algorithm.}
	\label{fig:more-mnist-reconstr_l2}
\end{figure*}

\begin{figure*}
	\begin{center}
		\begin{subfigure}[]{1\textwidth}
			\includegraphics[width=\textwidth]{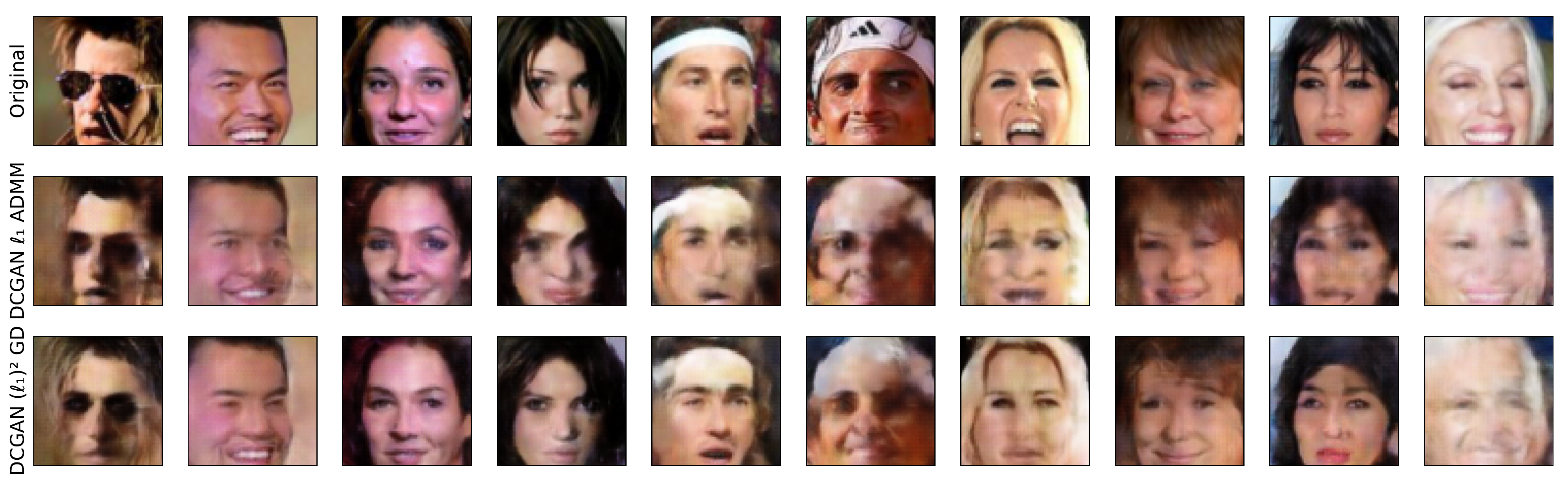}
			\caption{1000 measurements}
		\end{subfigure}
		\begin{subfigure}[]{1\textwidth}
			\includegraphics[width=\textwidth]{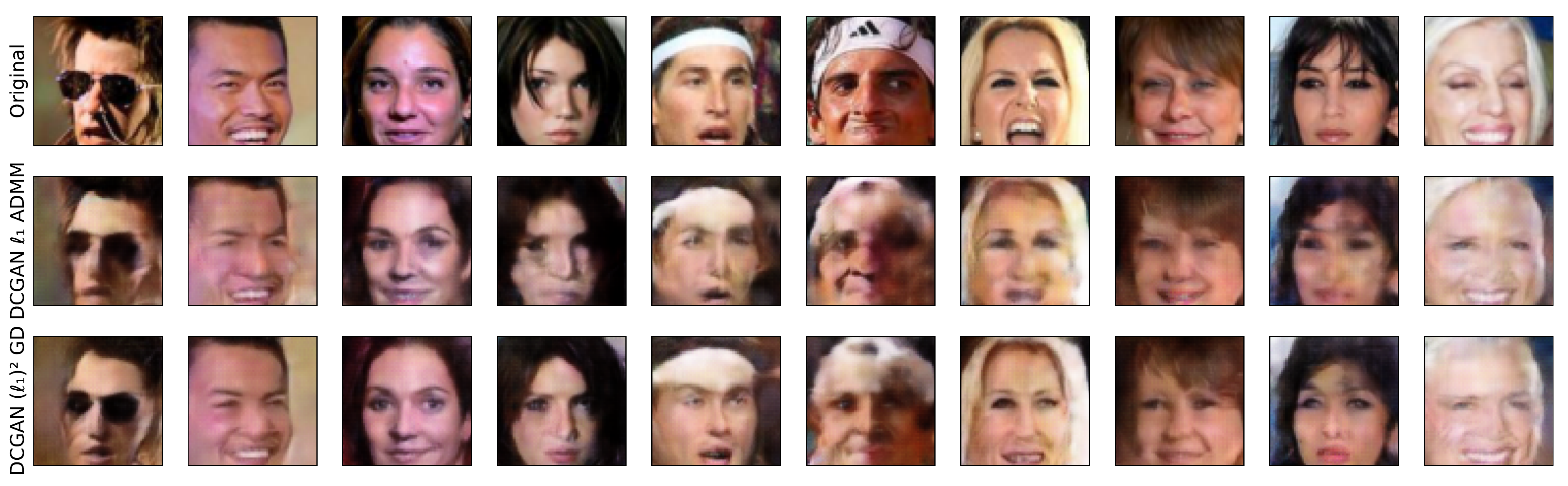}
			\caption{5000 measurements}
		\end{subfigure}
		\begin{subfigure}[]{1\textwidth}
			\includegraphics[width=\textwidth]{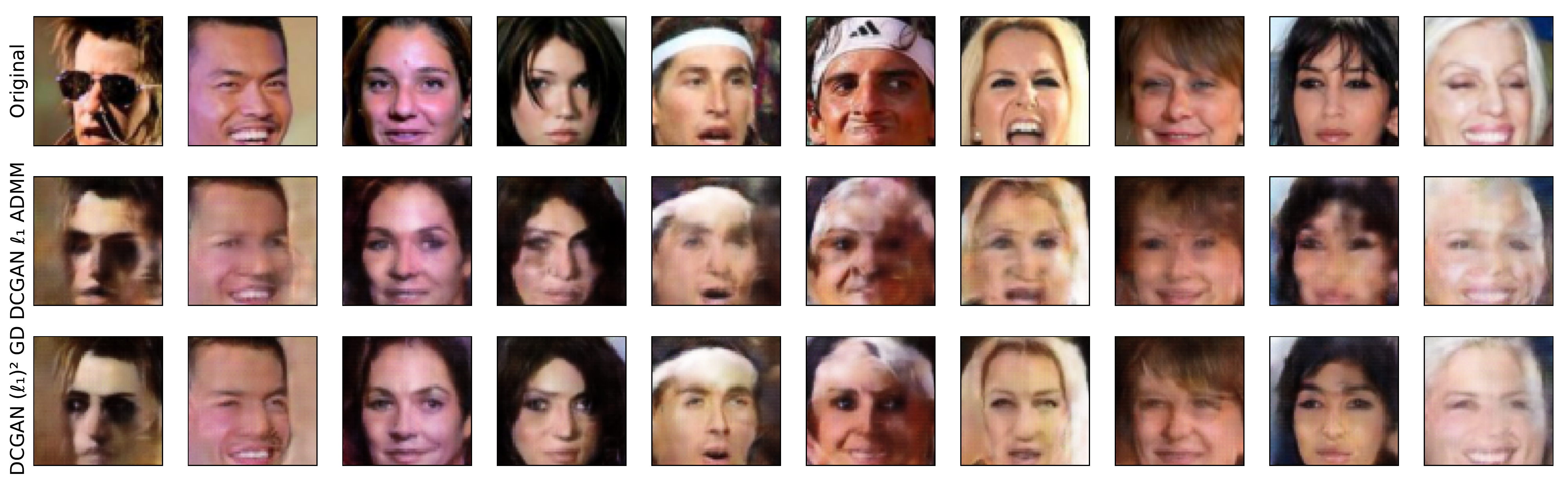}
			\caption{10000 measurements}
		\end{subfigure}
		\caption{Reconstruction results of CelebA samples with different numbers of measurements when the number of outliers is 3. Note that in each result, top row: original images, middle row: reconstructions by DCGAN with $\ell_1$ ADMM algorithm, and bottom row: reconstructions by DCGAN with $(\ell_1)^2$ GD algorithm.}
		\label{fig:more-celebA-reconstr1_l1}
	\end{center}
\end{figure*}

\begin{figure*}
	\begin{center}
		\begin{subfigure}[]{1\textwidth}
			\includegraphics[width=\textwidth]{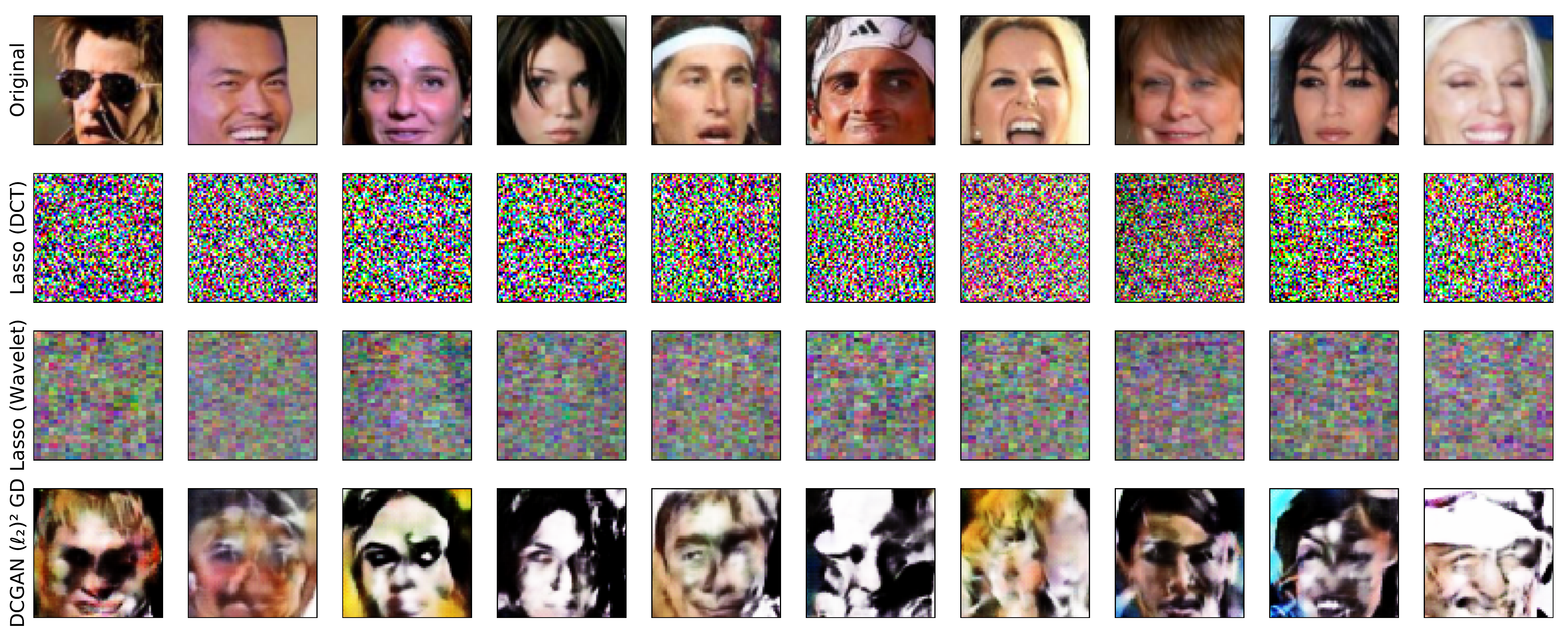}
			\caption{1000 measurements}
		\end{subfigure}
		\begin{subfigure}[]{1\textwidth}
			\includegraphics[width=\textwidth]{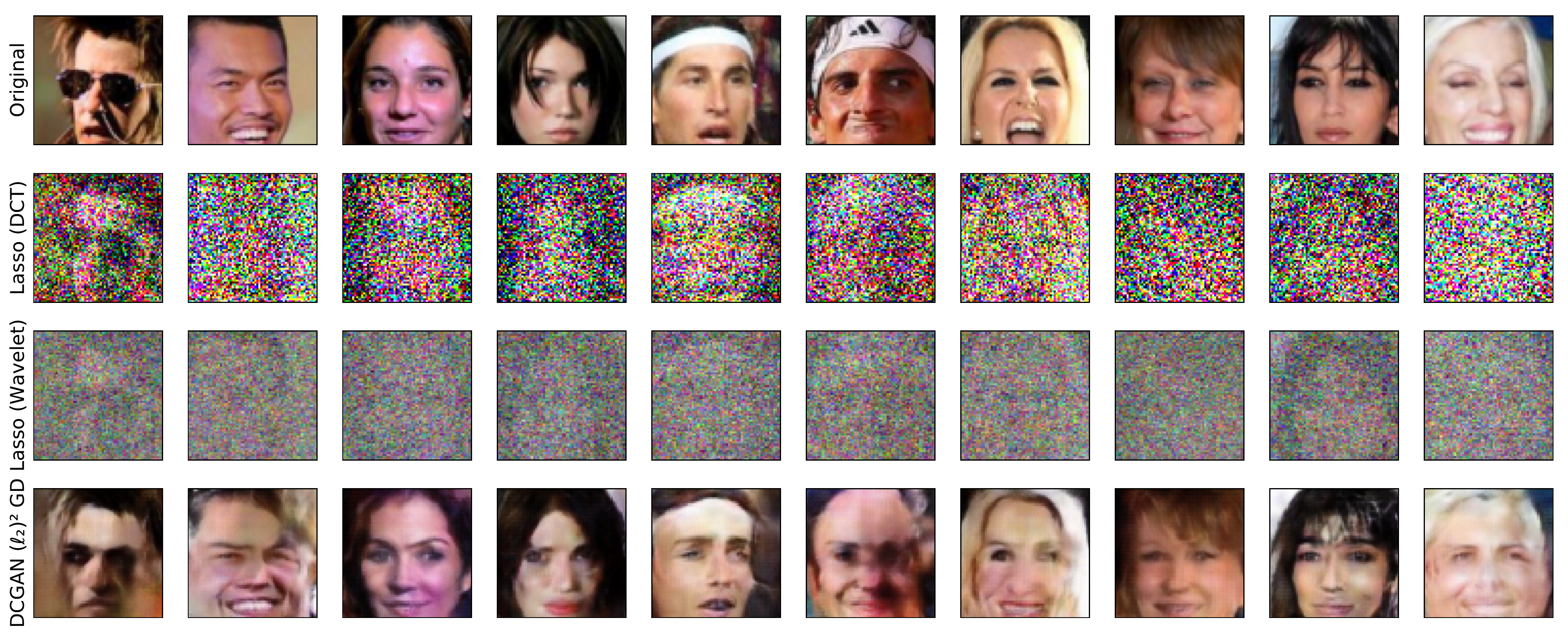}
			\caption{5000 measurements}
		\end{subfigure}
		\begin{subfigure}[]{1\textwidth}
			\includegraphics[width=\textwidth]{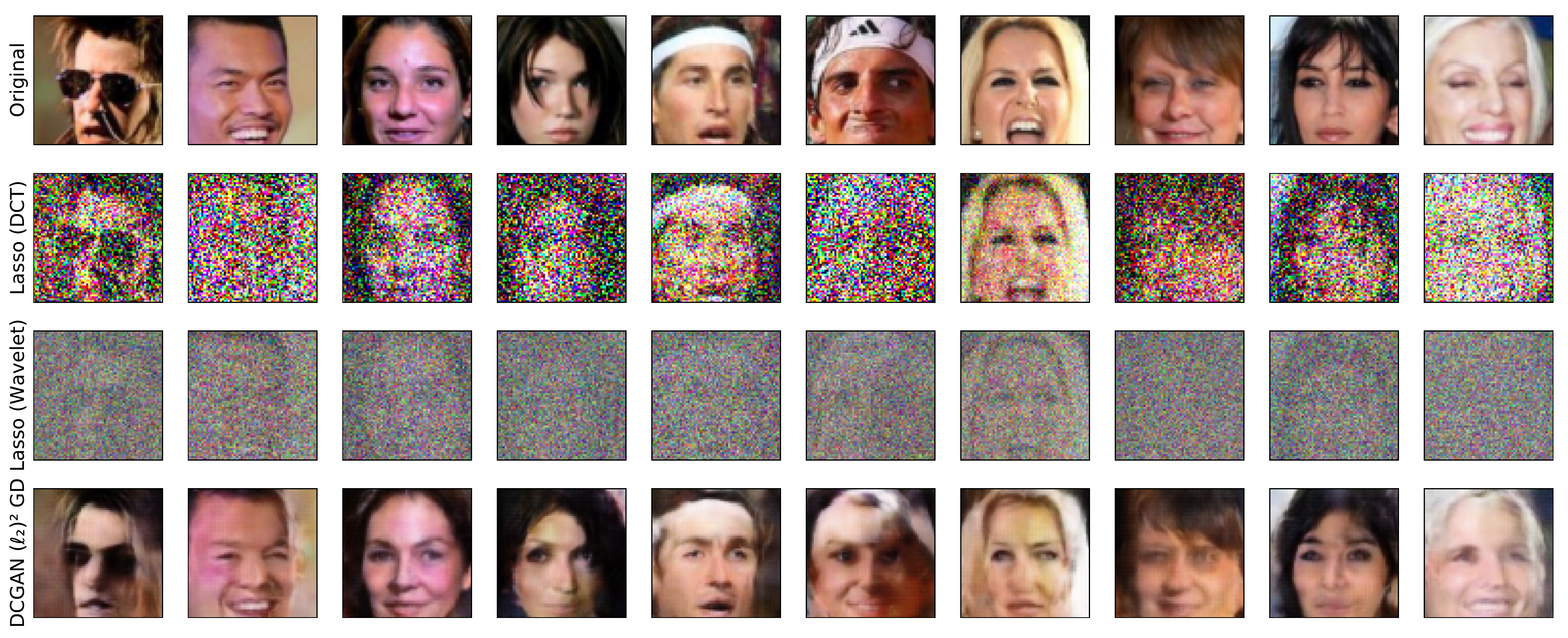}
			\caption{10000 measurements}
		\end{subfigure}
		\caption{Reconstruction results of CelebA samples with different numbers of measurements when the number of outliers is 3 (continue). In each row of results, top row: original images, second row: reconstructions by Lasso with DCT basis, third row: reconstructions by Lasso with wavelet basis, and last row: reconstructions by DCGAN with $(\ell_2)^2$ GD algorithm.}
		\label{fig:more-celebA-reconstr1_l2}
	\end{center}
\end{figure*}

\begin{figure*}
	\begin{subfigure}[t]{0.5\textwidth}
		\includegraphics[width=\textwidth]{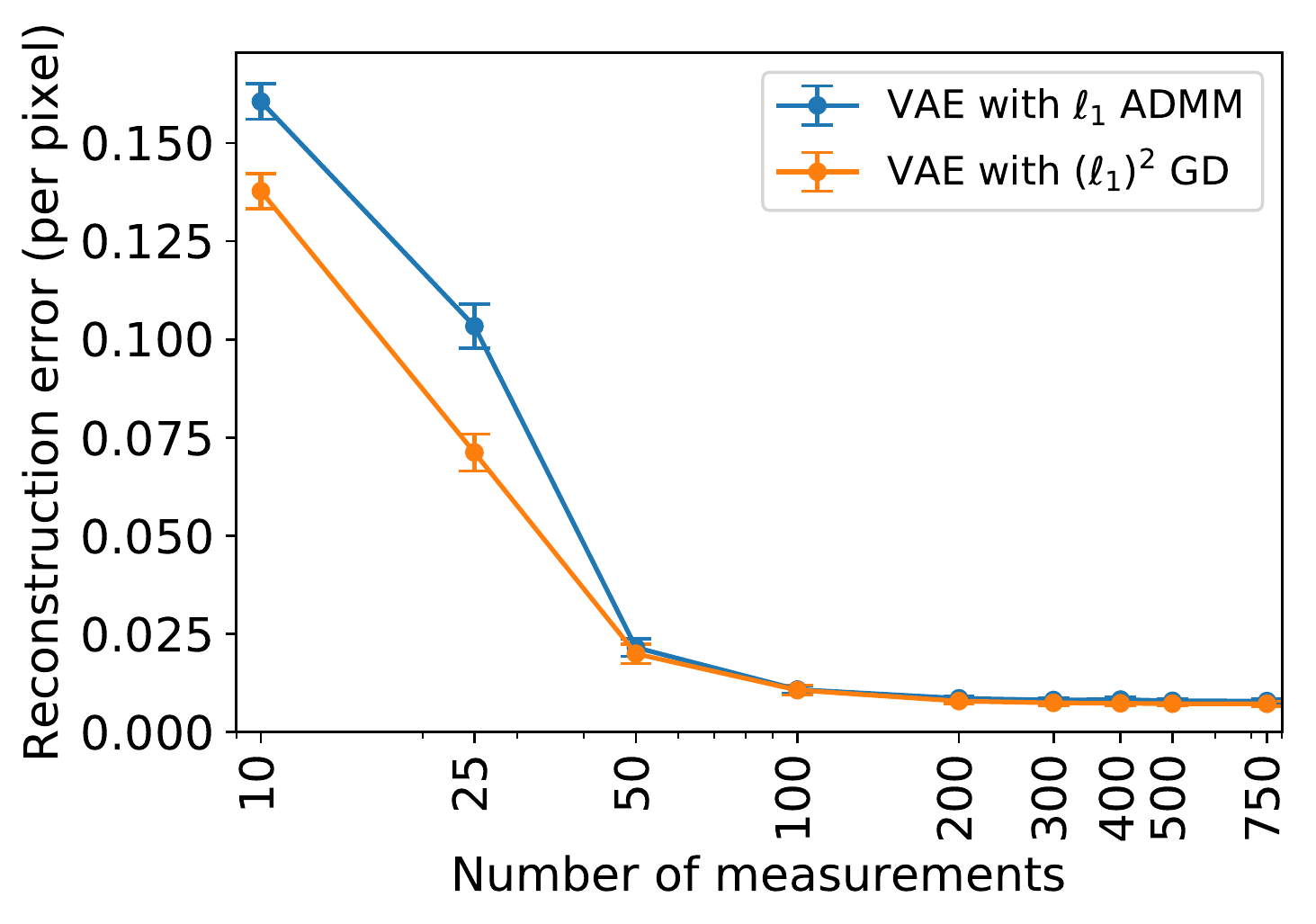}
		\caption{5 outliers}
	\end{subfigure}\hfill%
	\begin{subfigure}[t]{0.5\textwidth}
		\includegraphics[width=\textwidth]{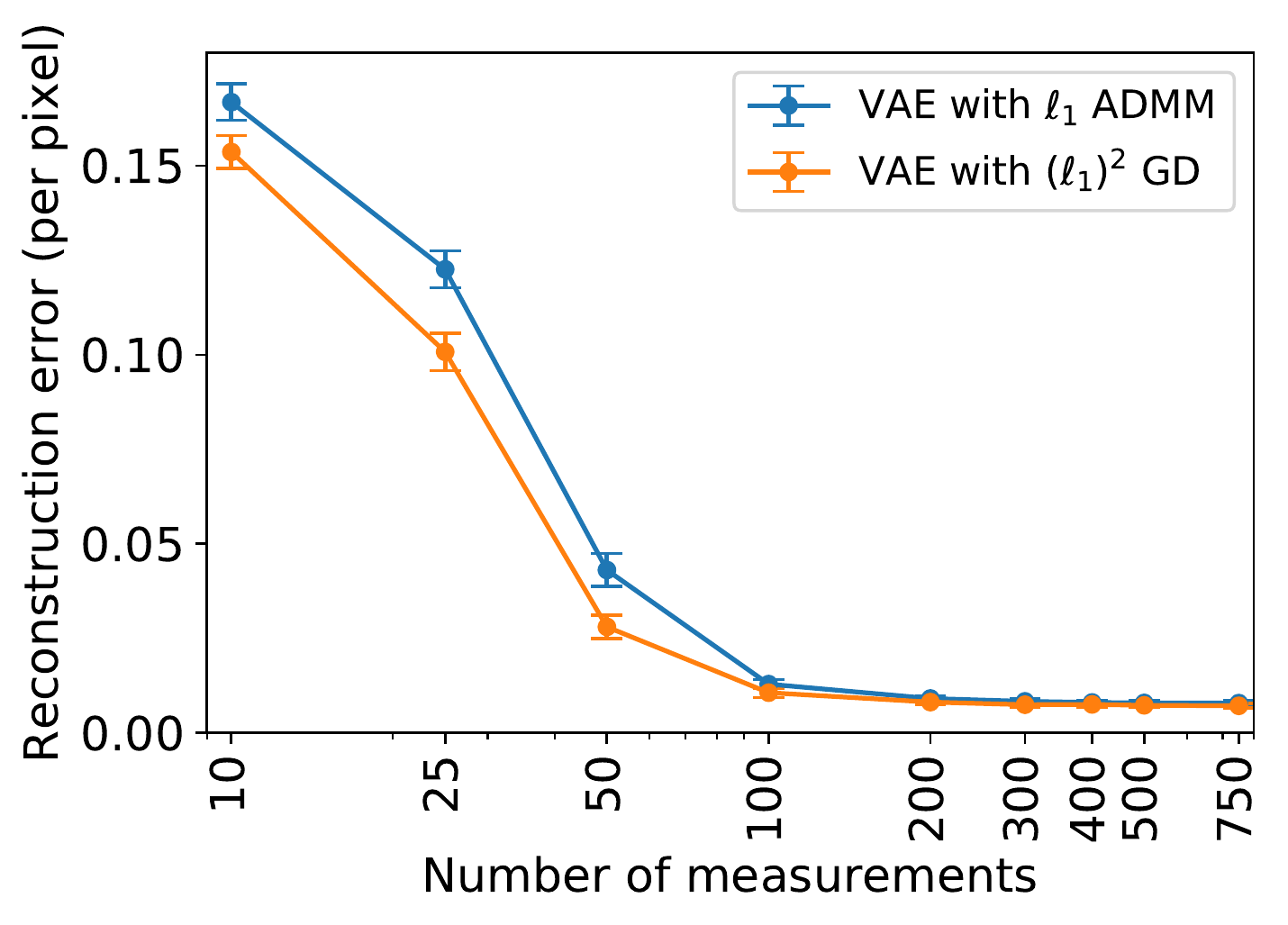}
		\caption{10 outliers}
	\end{subfigure}
	
	\begin{subfigure}[t]{0.5\textwidth}
		\includegraphics[width=\textwidth]{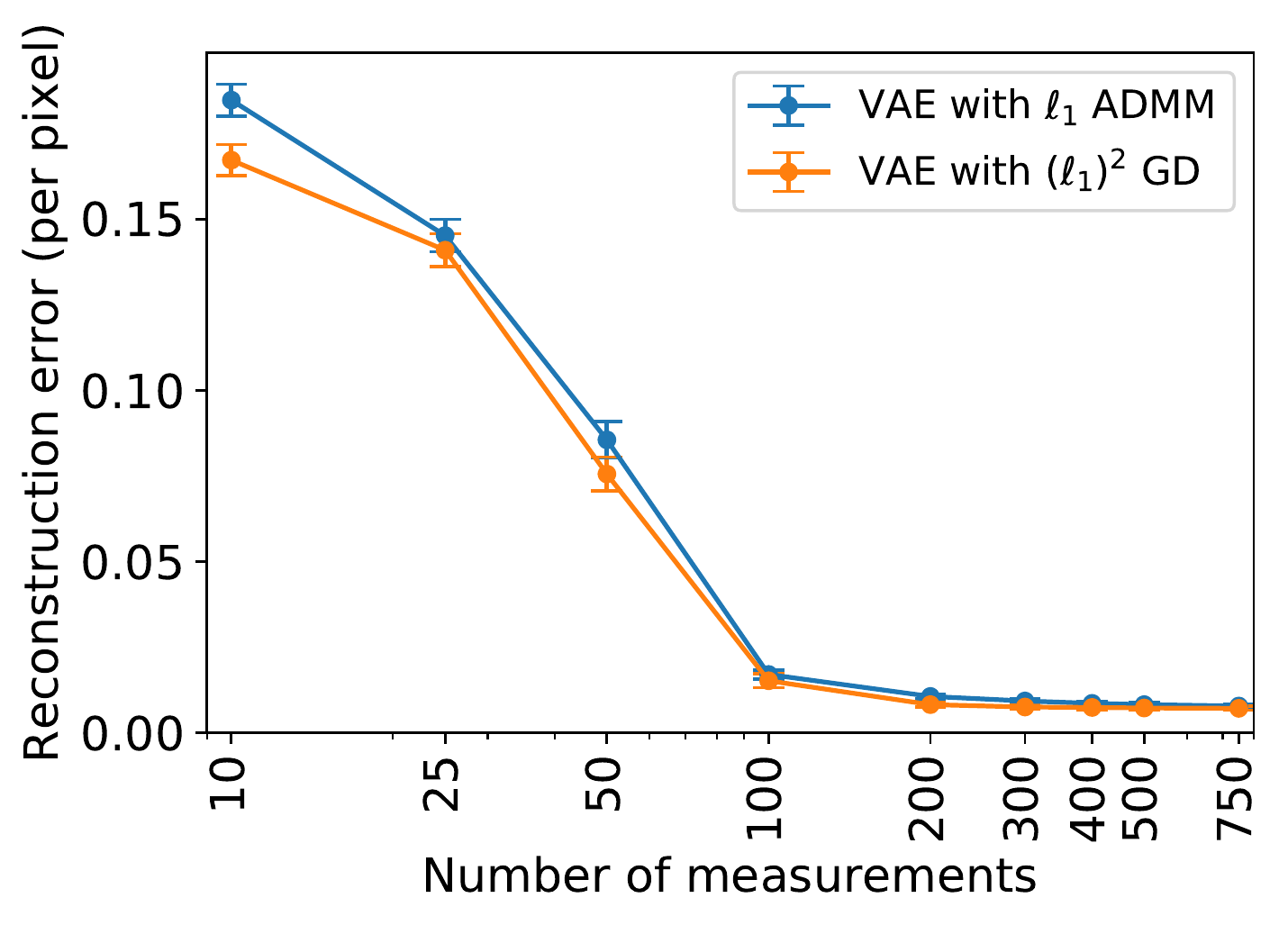}
		\caption{25 outliers}
	\end{subfigure}\hfill%
	\begin{subfigure}[t]{0.5\textwidth}
		\includegraphics[width=\textwidth]{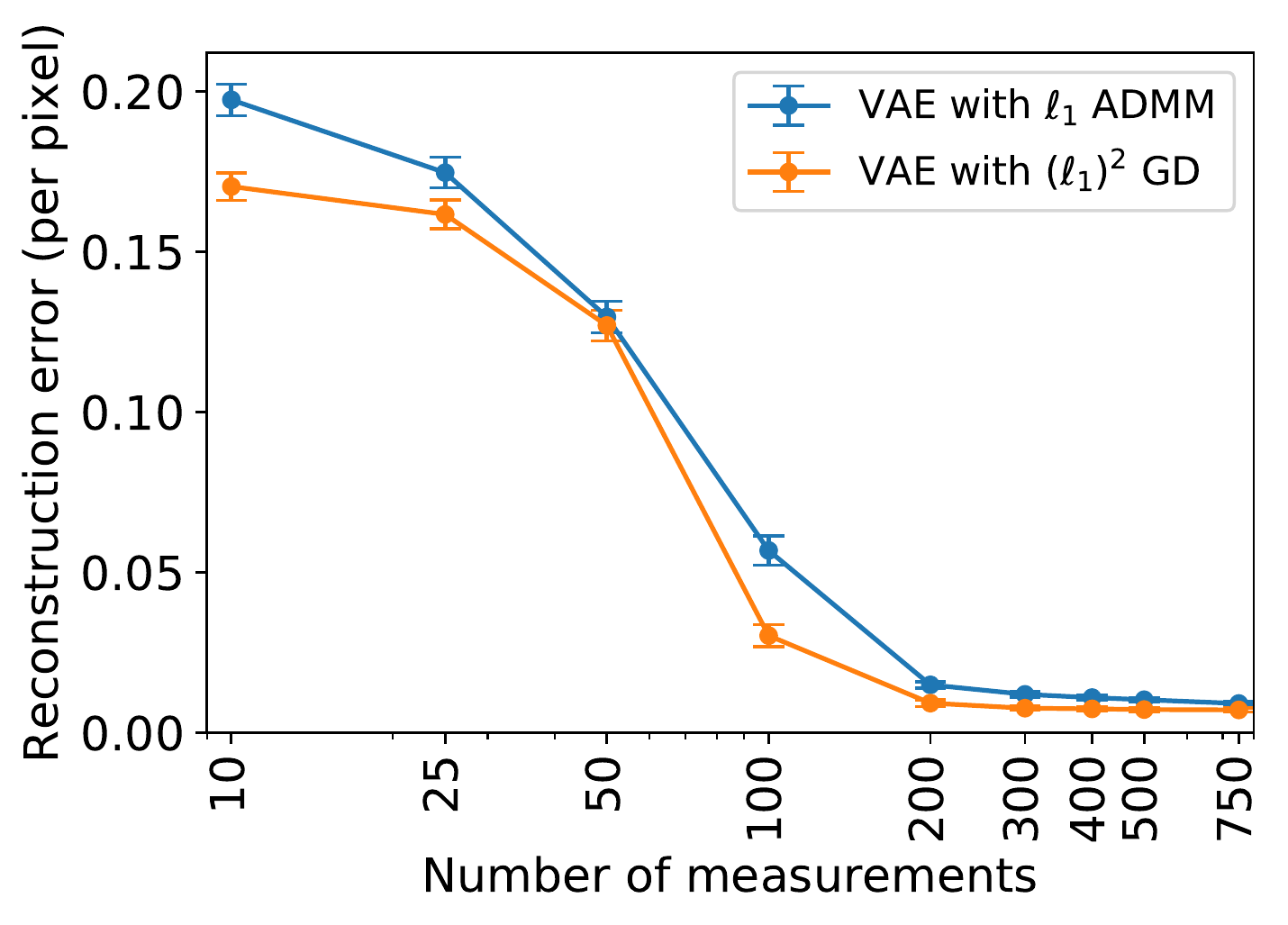}
		\caption{50 outliers}
	\end{subfigure}
	\caption{MNIST results: We compare the reconstruction error performance for VAE with $\ell_1$ ADMM and $(\ell_1)^2$ GD algorithms with different numbers of outliers.}
	\label{fig:mnist-reconstr-o5to50}
	\vspace*{2mm}
	
\end{figure*}

\begin{figure*}
	\begin{subfigure}[t]{0.5\textwidth}
		\includegraphics[width=\textwidth]{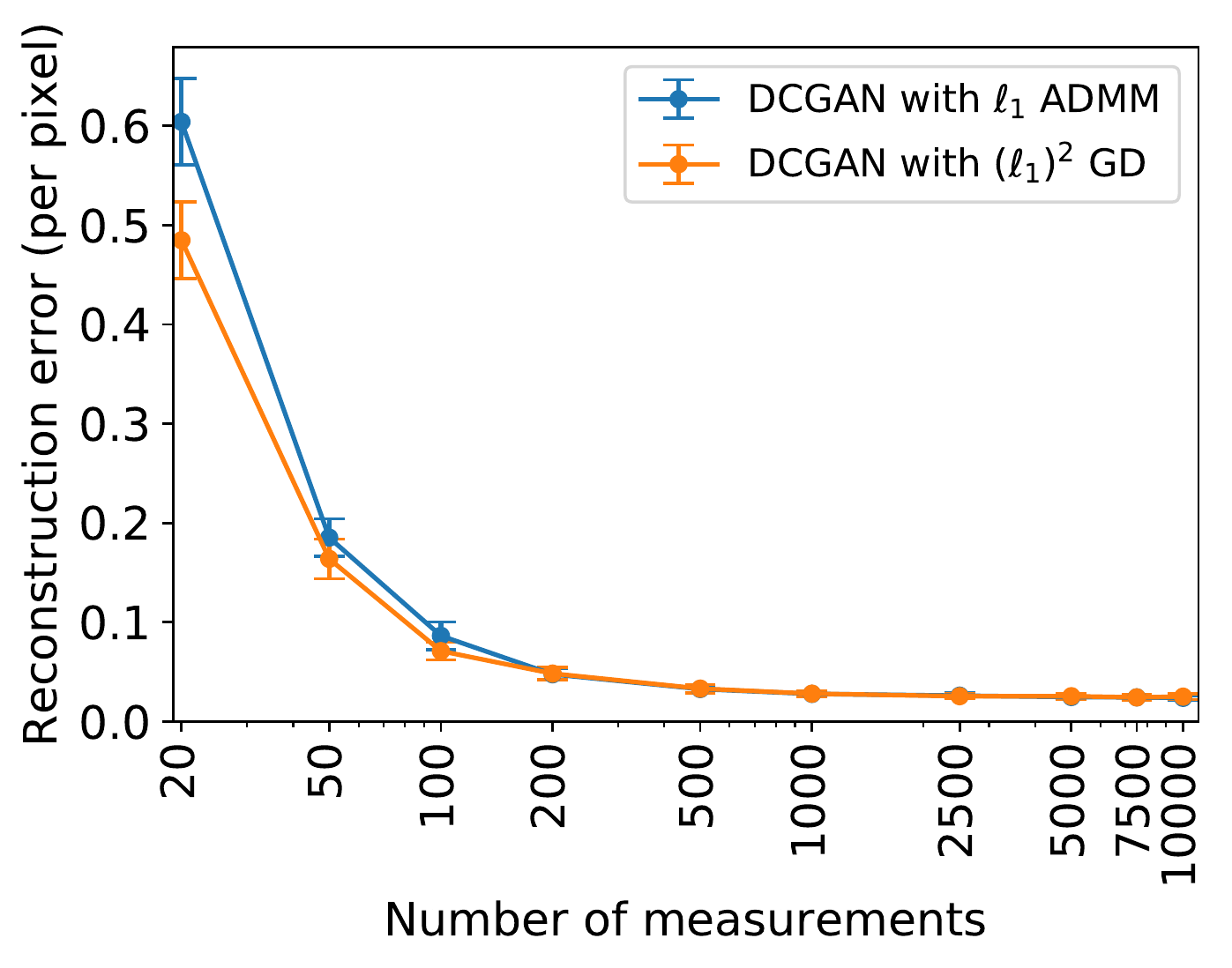}
		\caption{5 outliers}
	\end{subfigure}\hfill%
	\begin{subfigure}[t]{0.5\textwidth}
		\includegraphics[width=\textwidth]{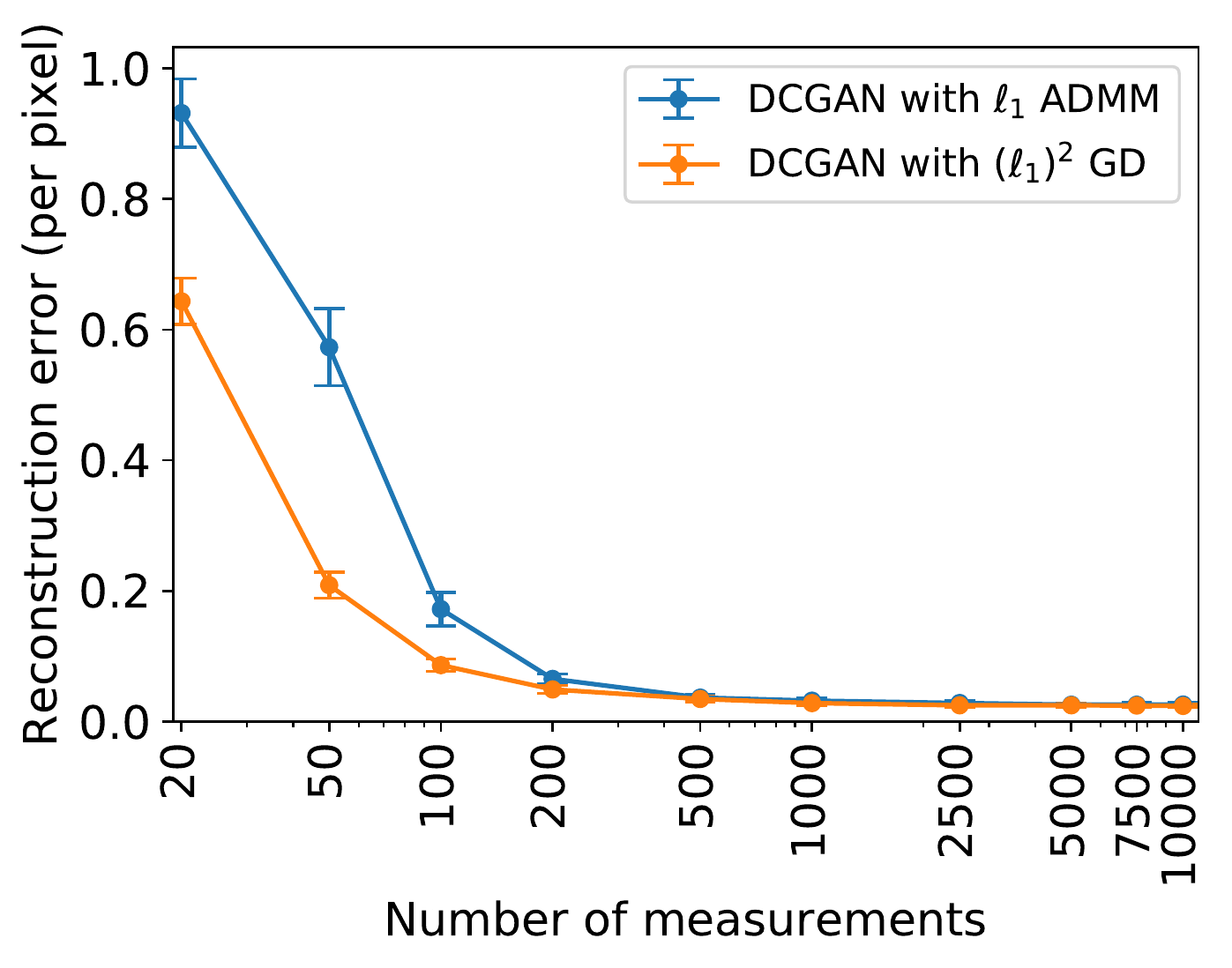}
		\caption{10 outliers}
	\end{subfigure}
	
	\begin{subfigure}[t]{0.5\textwidth}
		\includegraphics[width=\textwidth]{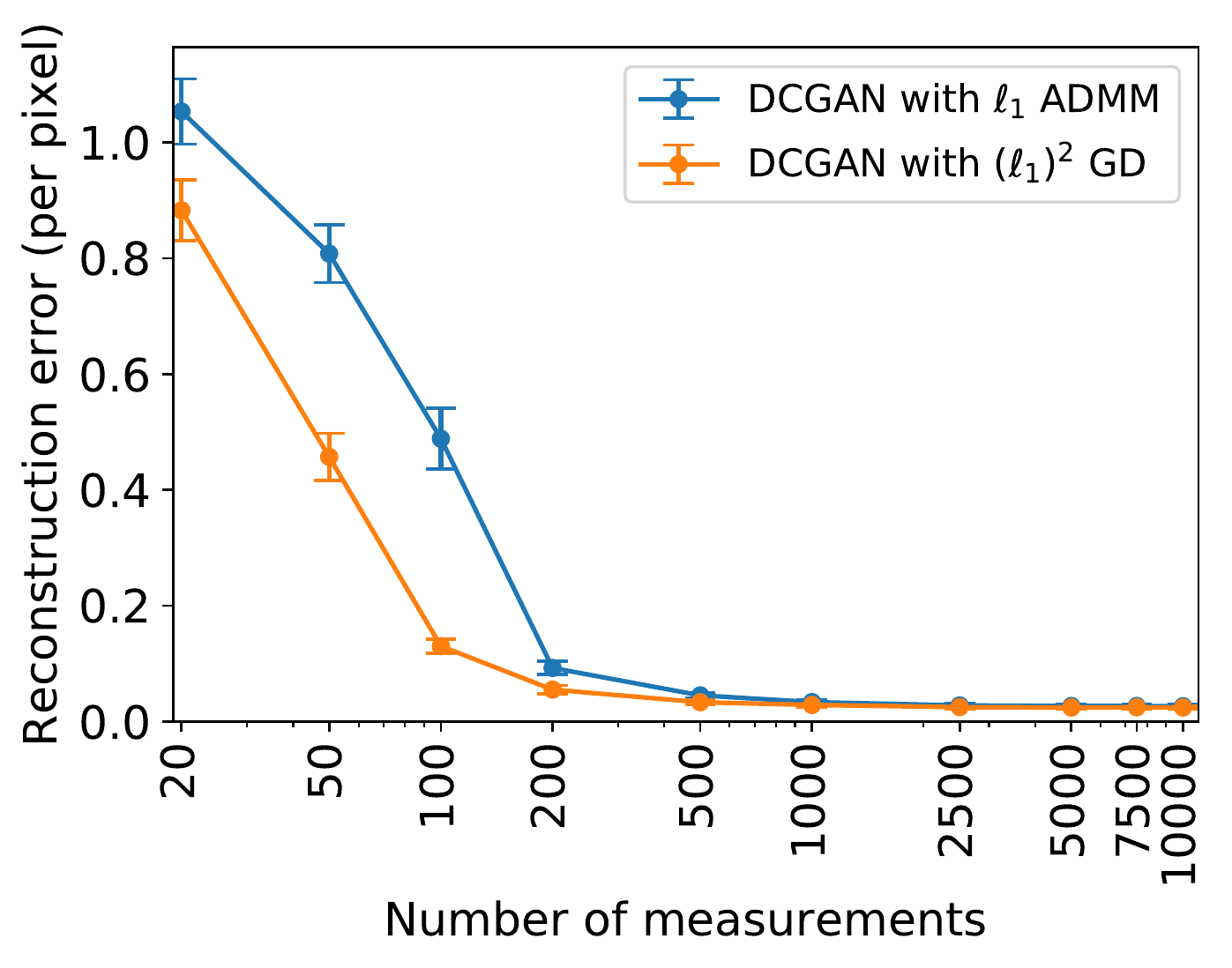}
		\caption{25 outliers}
	\end{subfigure}\hfill%
	\begin{subfigure}[t]{0.5\textwidth}
		\includegraphics[width=\textwidth]{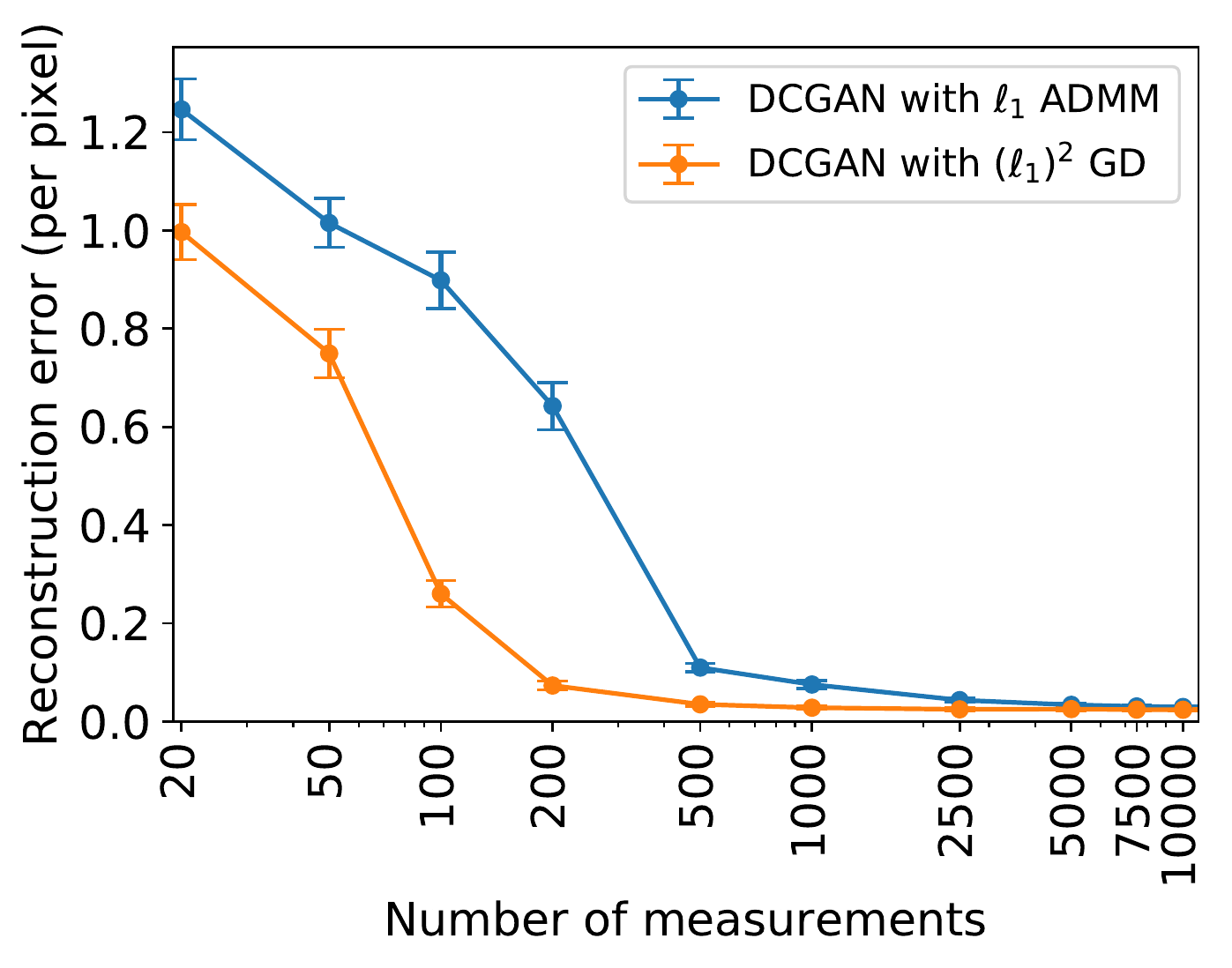}
		\caption{50 outliers}
	\end{subfigure}
	\caption{CelebA results: We compare the the reconstruction error performance for DCGAN with $\ell_1$ ADMM and $(\ell_1)^2$ GD algorithms with different numbers of outliers.}
	\label{fig:celebA-reconstr-o5to50}
	\vspace*{2mm}
	
\end{figure*}

\begin{figure*}
	\centering
	\begin{subfigure}[]{1\textwidth}
		\includegraphics[width=\textwidth]{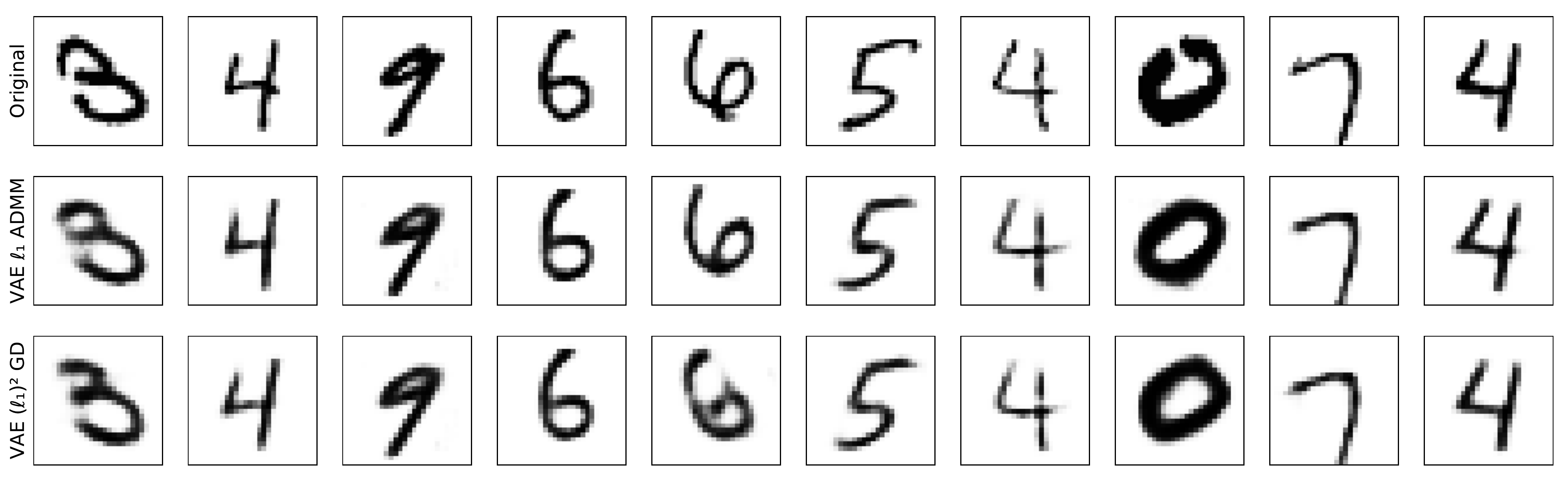}
		\caption{5 outliers}
	\end{subfigure}
	\begin{subfigure}[]{1\textwidth}
		\includegraphics[width=\textwidth]{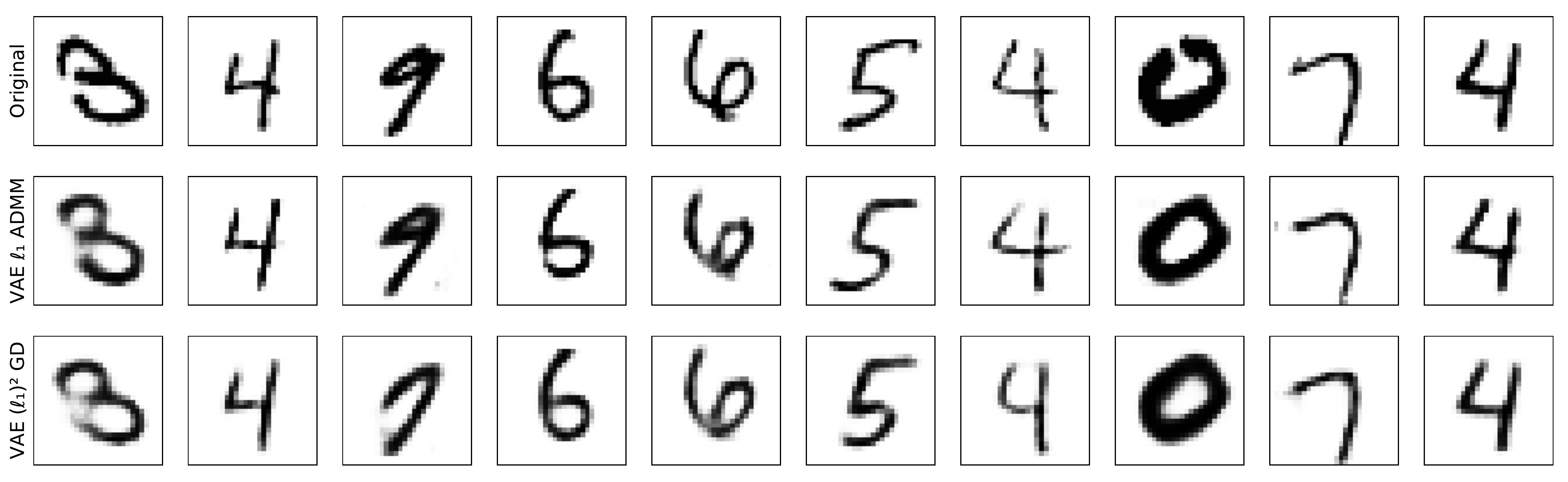}
		\caption{10 outliers}
	\end{subfigure}
	\begin{subfigure}[]{1\textwidth}
		\includegraphics[width=\textwidth]{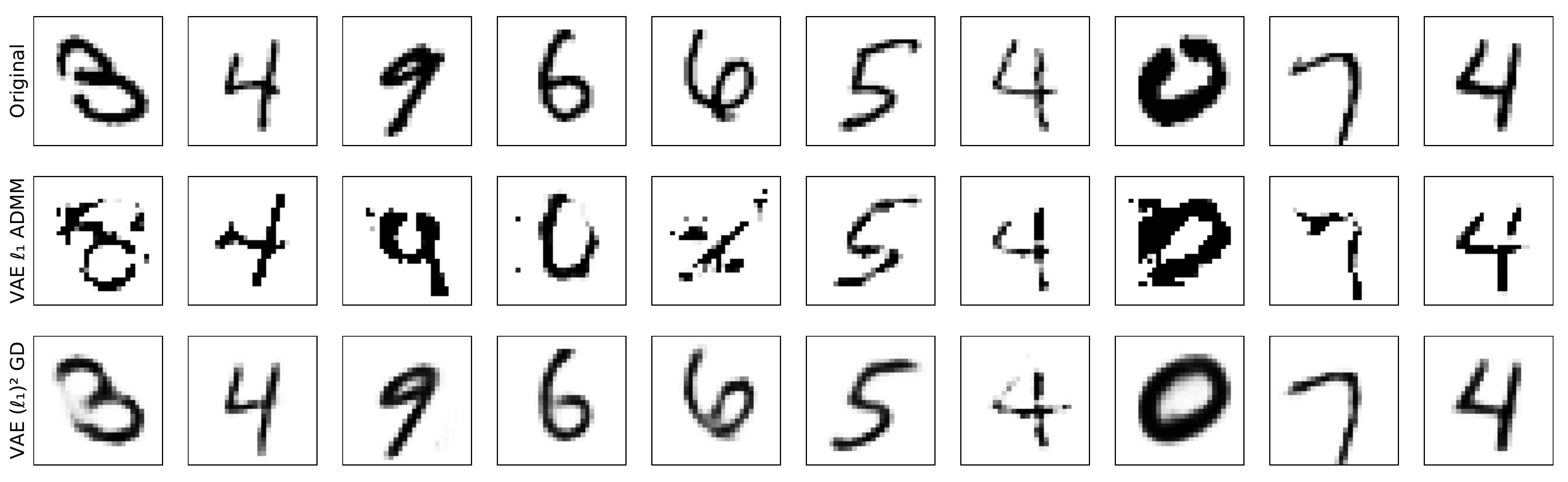}
		\caption{25 outliers}
	\end{subfigure}
	\caption{Reconstruction results of MNIST samples with different numbers of outliers when the number of measurements is 100. In each set of results, top row: original images, middle row: reconstructions by VAE with $\ell_1$ ADMM algorithm, and bottom row: reconstructions by VAE with $(\ell_1)^2$ GD algorithm.}
	\label{fig:mnist-sample-o1to25}
\end{figure*}

\begin{figure*}
	\begin{center}
		\begin{subfigure}[]{1\textwidth}
			\includegraphics[width=\textwidth]{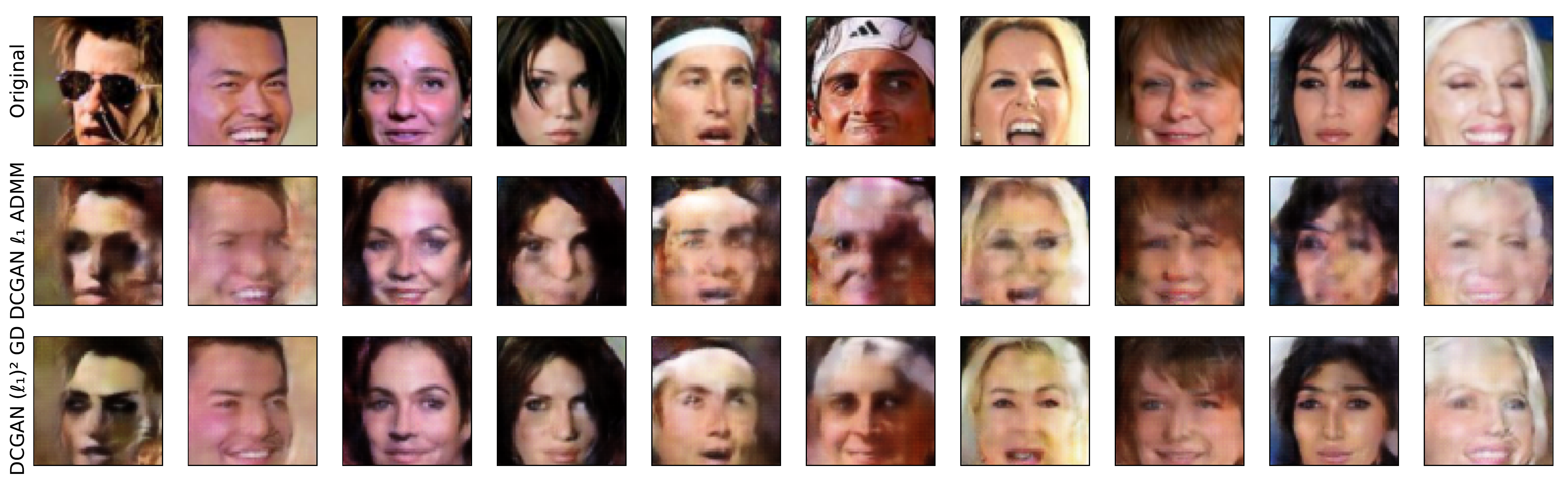}
			\caption{5 outliers}
		\end{subfigure}
		\begin{subfigure}[]{1\textwidth}
			\includegraphics[width=\textwidth]{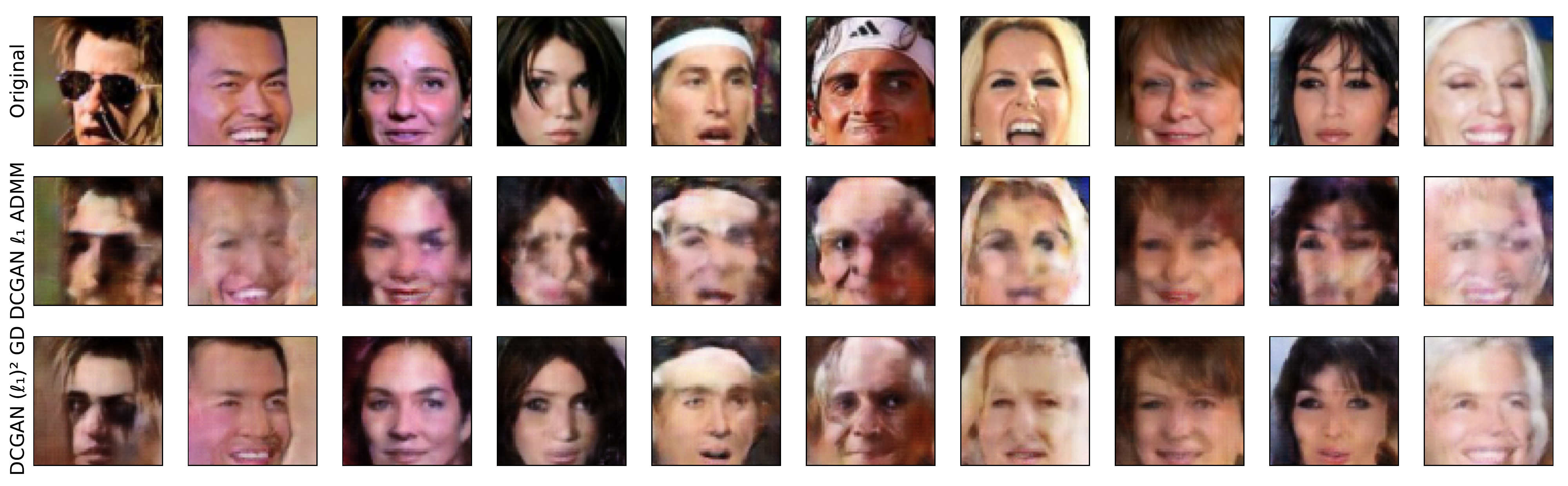}
			\caption{10 outliers}
		\end{subfigure}
		\begin{subfigure}[]{1\textwidth}
			\includegraphics[width=\textwidth]{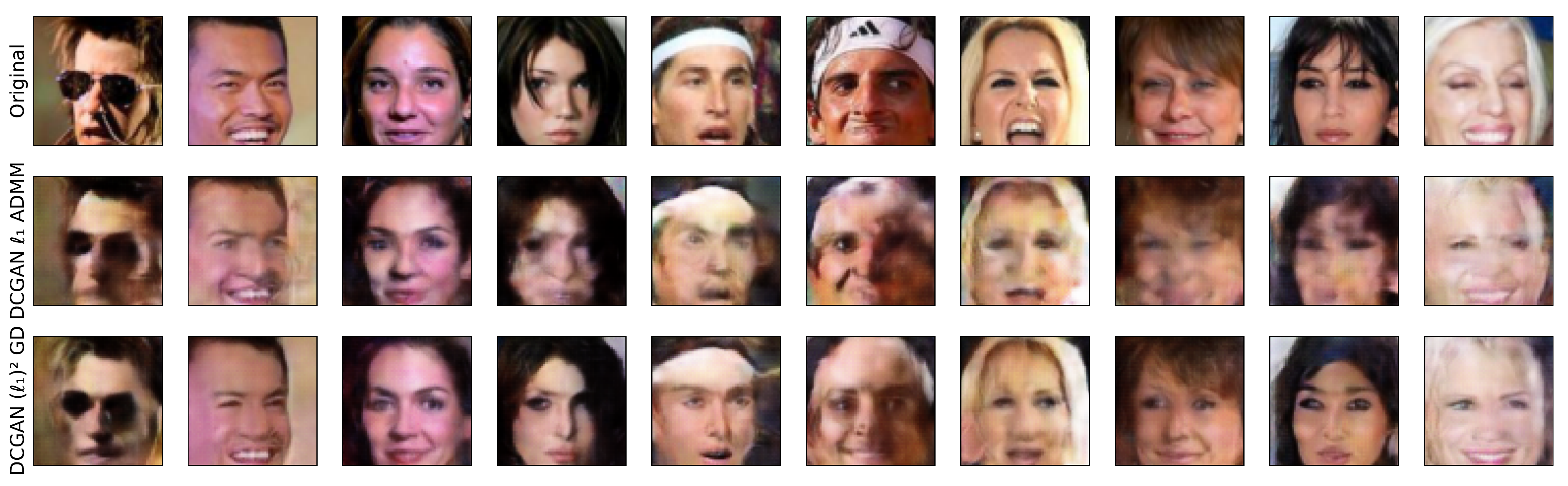}
			\caption{25 outliers}
		\end{subfigure}
		\begin{subfigure}[]{1\textwidth}
			\includegraphics[width=\textwidth]{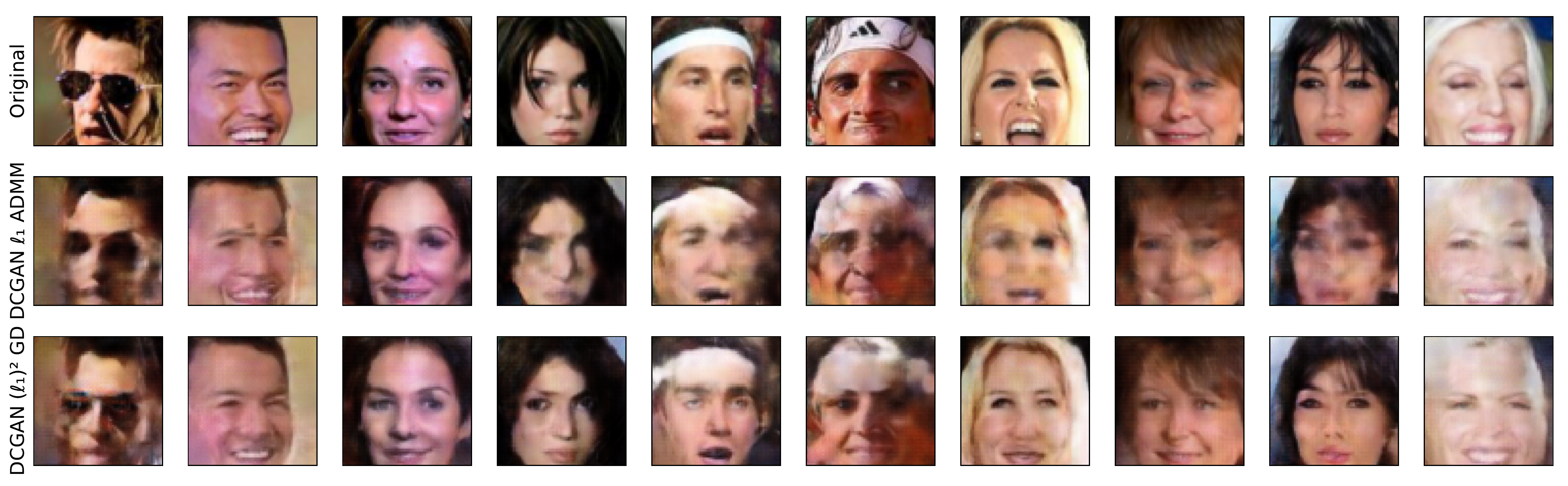}
			\caption{50 outliers}
		\end{subfigure}
		\caption{Reconstruction results of CelebA samples with different numbers of outliers when the number of measurements is 1000. In each set of results, top row: original images, middle row: reconstructions by DCGAN with $\ell_1$ ADMM algorithm, and bottom row: reconstructions by DCGAN with $(\ell_1)^2$ GD algorithm.}
		\label{fig:celebA-sample-o1to25-a}
	\end{center}
\end{figure*}

\end{document}